\documentclass[11pt,a4paper]{article}

\usepackage[margin=2.5cm]{geometry}
\usepackage{amsmath,amssymb,amsthm}
\usepackage{mathtools}
\usepackage{physics}
\usepackage{multirow}
\usepackage{jheppub} 
\hypersetup{colorlinks=true,
            linkcolor=blue!60!black,
            citecolor=blue!60!black,
            urlcolor=blue!60!black}
\usepackage{enumitem}
\usepackage{microtype}
\usepackage{booktabs}
\usepackage{array}
\usepackage{xcolor}
\usepackage{float}
\usepackage{tikz}
\usepackage[compat=1.1.0]{tikz-feynman}
\usepackage{graphicx}
\usepackage{subcaption}
\usepackage[T1]{fontenc}
\usepackage{lmodern}
\usepackage{microtype}
\usepackage{amsmath,amssymb,mathtools,bm,amsthm,mathrsfs}
\usepackage{booktabs,tabularx,array,multirow}
\usepackage[dvipsnames]{xcolor}
\usetikzlibrary{arrows.meta,calc,positioning}
\usepackage[most]{tcolorbox}
\newcommand{\U}{\mathrm{U}}
\newcommand{\Com}{\mathrm{Com}}

\newcommand{\eps}{\varepsilon}

\newcommand{\Lag}{\mathcal{L}}
\newcommand{\A}{\mathcal{A}}
\newcommand{\pr}{\prec}

\newcommand{\cO}{\mathcal{O}}

\newcommand{\nw}[1]{#1}

\begin{document}
\title{\boldmath Fermion Mass and Mixing Patterns from Fusion Algebras}

\author{Jacques Distler,$^a$ Can Kilic,$^a$  Sanjay Mathai$^{a}$}

\affiliation{$^a$ Theory Group, Weinberg Institute for Theoretical Physics,\\ The University of Texas at Austin, Austin TX  78752}

\emailAdd{distler@physics.utexas.edu}
\emailAdd{kilic@physics.utexas.edu}
\emailAdd{sanjaym@utexas.edu}

\abstract{
 The large hierarchy of fermion masses in the Standard Model has long posed a challenge to be explained. One approach  is to try to invoke a symmetry principle to dictate the structure (``texture'') of the matrices of fermion Yukawa couplings to the Higgs field. Here we invoke \emph{noninvertible} symmetries, specifically, fusion algebras, to provide such textures. In particular, the structure of the fusion-algebra can be used to assign labels to each field, which determines how suppressed any given composite operator is as a term in the Lagrangian. In UV completions, this can be interpreted as specifying a hierarchy of operators that in the IR effective theory become the Yukawa couplings to the Higgs field. Remaining agnostic of a specific UV model, we perform a Monte Carlo numerical scan over unknown Wilson coefficients, and we identify the rank-8 fusion algebra $\rm{FR}^{8,6}_8$ as the one providing the best fit for the textures of the up-quark, down-quark and charged-lepton Yukawa matrices. We also study the predictions for the CKM and PMNS mixing matrices, with the conclusion that none of the fusion algebras we study provide an accurate fit for the masses and mixing angles simultaneously. The fusion algebra structure ensures that the textures that we find are stable under RG running and persist down to low energies.
}

\preprint{UT-WI-36-2026}

\maketitle

% ============================================================
\section{Introduction}
\label{sec:intro}
% ============================================================

The Standard Model (SM) offers no explanation for the broad spread of
fermion masses among the three generations.  According to the PDG \cite{ParticleDataGroup:2024cfk}, at the scale $M_Z$, the ratios of charged lepton masses stand at $m_e:m_\mu:m_\tau\approx1:210.7:3578$, the down-type quarks at
$1:19.9:1058$, and the up-type quarks at $1:506:1.374\times10^{5}$. In the SM these numbers
are free parameters, which could all have been of the same order of magnitude. 

The classic answer to this puzzle is to impose a flavor symmetry, weakly broken by one or several spurions.
Froggatt and Nielsen introduced the first systematic
mechanism of this type~\cite{Froggatt:1978nt}, in which a spontaneously broken Abelian flavor
symmetry $\U(1)_{\rm FN}$ and a ``flavon'' field $\phi$ generate each Yukawa entry at
a power of $\langle\phi\rangle/\Lambda$ fixed by the charges. Non-Abelian discrete flavor
symmetries such as $A_4$, $S_4$ and $\Delta(27)$ gained popularity in the
neutrino era \cite{Altarelli:2010gt,Ishimori:2010au}, and more recently
modular invariance has been made to play the same role
\cite{Feruglio:2017spp}.  Further ideas where the mechanisms that do not rest on a symmetry at all,
such as wave-function localisation in extra dimensions
\cite{Arkani-Hamed:1999ylh} or partial compositeness \cite{Kaplan:1991dc},
generate the hierarchy from exponentially small overlaps or from anomalous
dimensions instead.  

Over the past decade it has become clear that quantum field theories
generically admit symmetries that are not groups at all
\cite{Gaiotto:2014kfa}.  They are implemented by topological defect operators which fuse with
non-group-like rules, so that the defects need not be invertible and the
fusion is governed by a fusion category rather than by a group
\cite{McGreevy:2022oyu,Cordova:2022ruw,Schafer-Nameki:2023jdn,Bhardwaj:2023kri,Shao:2023gho,Kaidi:2026urc}.
Explicit non-invertible duality defects in four dimensions were constructed in
refs.~\cite{Kaidi:2021xfk,Choi:2021kmx}. In perturbative string theory,
worldsheet non-invertible symmetries govern selection rules on spacetime
amplitudes through the loop-order structure of genus-$g$ Riemann surfaces.
Such symmetries have since been shown to act on the Standard Model itself
\cite{Choi:2022jqy,Cordova:2022qtz}, to control the smallness of neutrino
masses \cite{Cordova:2022fhg}, and more generally to generate exponential
hierarchies \cite{Cordova:2022ieu}.

The algebraic structure that forms the backbone of such a symmetry is called a fusion algebra $\A$, a finite-dimensional algebra with a preferred basis of ``simple elements'' whose structure constants for their products are non-negative integers.  The feature that makes a fusion algebra useful for flavor is
that, unlike a group, its selection rules are graded. An operator forbidden at leading order can become allowed once multiple ``commutator'' insertions are made, and the number of insertions needed is fixed by the fusion algebra. A suppression hierarchy is therefore built into the structure of the fusion algebra, and this hierarchy is independent of its UV origins, be they spurions or loops of heavy particles.

The fusion algebra, and its attendant selection rules are presumably the low-energy residue of a symmetry of the high-energy theory. In one realization, loops of the heavy fields generate the hierarchy of entries in the Yukawa matrix; in another, the same entries are generated by insertions of spurion vacuum expectation values. Regardless of the mechanism, after integrating out the heavy fields, we then need to run the Yukawa couplings down to low energies. We don't want the texture generated at high energies to be wiped out by the RG running. For the up Yukawa matrix, the 1-loop running coupling is
\begin{equation}
Y_u(E) = Y_u(\Lambda) + (Y_u Y^\dagger_u-Y_d Y_d^\dagger)Y_u  \frac{3}{32\pi^2} \log(E/\Lambda)+\dots
\end{equation}
where the ``$\dots$'' represent other contributions (e.g.~from gauge-boson exchange) which are subleading in this expansion. To avoid spoiling the texture, it suffices that
\begin{subequations}\label{RGstab}
\begin{equation}\label{RGstabUp}
(Y_u Y^\dagger_u-Y_dY^\dagger _d)Y_u \lesssim Y_u
\end{equation}
where by ``$\lesssim$'' we mean that the power of $\eps$ that appears in each entry of the matrix on the left is greater than or equal to the power of $\epsilon$ that appears in the corresponding matrix on the right. For the down quark and lepton Yukawa matrices, \eqref{RGstabUp} is replaced by
\begin{equation}
(Y_d Y_d^\dagger-Y_u Y_u^\dagger )Y_d \lesssim Y_d,\qquad Y_e Y_e^\dagger Y_e \lesssim Y_e
\end{equation}
 \end{subequations}
\eqref{RGstab} are nontrivial constraints on the form of the textures of the Yukawa matrices. In Appendix \ref{app:RGstab}, we show that they follow from the structure of the fusion  when the Higgs is assigned an invertible element of the fusion algebra.

A simple way to imbue a Lagrangian with a fusion algebra structure is to use conjugacy classes of a finite group $G$ to label the fields instead of irreducible representations of the group. This especially arises in twisted sectors of a $G-$orbifold where closed strings on the orbifold are labelled by conjugacy classes of the group $G$ \cite{Heckman:2024obe,Kaidi:2024wio,Kobayashi:2025ocp}. In order to effectively explain the texture of masses within the three generations of the SM, we need groups whose conjugacy classes closes at least after two steps. This severely restricts the possible candidate groups. Fusion algebras generalise this construction, and the phenomenology of the resulting Yukawa textures has begun to be explored in refs.~\cite{Kobayashi:2025cwx,Kobayashi:2025znw,Kobayashi:2025lar,Suzuki:2025bxg,Suzuki:2025kxz,Dong:2025jra,Kobayashi:2025wty,Dong:2026crl,Kobayashi:2026msq,Ding:2026sfm, Otsuka:2026fpy}. 

In this paper, we set up the fusion-algebra selection rule for Yukawa operators and make explicit the one structural concept that organises everything that follows. With the entries of the Yukawa matrices drawn from a geometric hierarchy $\{1,\eps,\eps^2,\dots , \eps^{\rm{cl}(\A) } \}$, the largest attainable singular value spread is of the order $\eps^{\rm{cl}(\A) }$ for generic $\cO (1)$ coefficients, where $\rm{cl}(\A) $ is the commutator length of the algebra $\A$ (to be defined in the next section). Applied to the SM, the up-type quark sector immediately restricts the candidate fusion algebras to be at least of $\rm{cl}(\A) =3$. Therefore, we catalogue all multiplicity-free fusion algebras of rank up to $10$ along with their commutator length and run a scan over all possible textures attainable using these algebras. We specifically find that there are exactly $4$ multiplicity-free $\rm{cl}(\A) =3$ fusion algebras of rank$\leq 10$ and assign scores to the best fit assignment for each algebra after running a Monte-Carlo scan over the undetermined $\cO(1)$ coefficients. We find that there exists one fusion algebra of rank 8, and  3 at rank 10, that can reproduce the mass hierarchies in all three fermion sectors. We also investigate whether mixing patterns in the quark and neutrino sectors can be fitted simultaneously with the masses, however we find no candidate algebra within the set defined above that achieves a high quality fit in this case.

The paper is organized as follows. In Sec.~\ref{sec:framework} we review fusion algebras and define the commutator filtration $\Com(\A)^L$ that sets the selection rules at each order. Sec.~\ref{sec:yukawa-setup} turns these rules into Yukawa textures for the leptons, quarks and neutrinos, and introduces the reach bound and the scoring metrics. In Sec.~\ref{sec:cl2} we show that algebras with $\rm{cl}(\A)=2$ that are of rank $\leq 7$ cannot generate the up-quark hierarchy. Sec.~\ref{sec:rank8} analyses the unique $\rm{cl}=3$ algebra of rank $\leq 9$, $\rm{FR}^{8,6}_8$, and its CKM matrix, and Sec.~\ref{sec:rank8-neutrino} its neutrino masses and PMNS matrix. We conclude in Sec.~\ref{sec:discussion} with discussion and open questions. The appendices collect the fusion tables in the labelling we have used, the detailed $\rm{cl}(\A)=2$ analysis, example diagrams, the rank-10 algebras, and the proof that the textures are RG stable (Appendix~\ref{app:RGstab}).

% ============================================================
\section{The Fusion-Algebra Framework}
\label{sec:framework}
% ============================================================

Let us begin by briefly describing what a fusion algebra is and how it
can be used as an organizational tool for effective field theories.  A \emph{fusion algebra} is
a finite-dimensional associative algebra $\A$ (a vector space over $\mathbb{C}$, endowed with an associative and distributive product) such that:
\begin{itemize}
\item There is a distinguished set $A$ of \emph{simple elements} which form a basis for $\A$.
\item The product on elements of $A$
\begin{equation}
    a\cdot b =\sum_{c\in A}N_{ab}^c c
\end{equation}
with $N_{ab}^c$ non-negative integers. We will exclusively consider commutative \emph{multiplicity-free} fusion algebras, for which $N_{ab}^c\in\{0,1\}$. We denote $c\prec a\cdot b$ whenever $N_{ab}^c=1$, ie. whenever $c$ appears in the product of $a$ with $b$.
\item The multiplicative identity $e$ of $\A$ is an element of $A$. Thus $N_{eb}^a=N_{be}^a=\delta^{a}_{b}$.
\item $A$ has ``duals.'' That is, for each $a\in A$, there is a unique $\overline{a}\in A$ such that $N_{a\overline{a}}^e=N_{\overline{a}a}^e=1$. 
\item \emph{Frobenius Reciprocity:} $N_{ab}^c=N_{\overline{a}c}^b= N_{c\overline{b}}^a,\; \forall a,b,c\in A$.
\end{itemize}

The number of basis elements $r=|A|$ is called
the \emph{rank} of the fusion algebra.  Fusion algebras of this kind arise as
the Grothendieck rings of fusion categories \cite{Etingof:2002vpd}, and contain
$\mathrm{Rep}(G)$, $\mathrm{Conj}(G)$ and the Tambara--Yamagami rings
\cite{Tambara:1998vmj} as special cases.

Now suppose we are given a theory with a set of fields
$\phi_1,\phi_2,\dots,\phi_n$, not necessarily scalars.  Similarly to assigning
representations of the gauge and global symmetries of the theory to such fields, if the theory
admits a fusion-algebra structure one assigns a basis element to each field.  We
denote such an assignment by $\phi[a]$ with $a\in A$, and the conjugate field then
carries $\phi^{*}[\bar{a}]$.

\paragraph{Selection rules from fusion algebras:}
Given an assignment of the basis elements to the respective fields, we allow a tree level vertex $\cO=\phi_1 \phi_2 \cdots \phi_n$ with fields $\phi_i[a_i]$ in the EFT lagrangian only if 
\begin{equation}
    e\prec a_1\cdot a_2\cdot  \ldots \cdot a_n
\end{equation}
i.e. the product $a_1\cdot a_2\cdots a_n$ must contain $e$ with
coefficient $1$.  This is the direct analogue of the way in which group
representations determine selection rules.

What is new here is that, unlike group representations --- whose
selection rules are the same at all loop orders --- the selection rules of a
fusion algebra can differ from one loop order to the next\footnote{Here we're using ``loop order'' to denote the order in the grading induced by the fusion algebra structure. That's intended to be merely suggestive; loops of heavy particles is one way to generate such a hierarchy, but we're not actually committing ourselves to any particular UV mechanism.}.
To make this precise we define the commutator sets $\Com (\A)^L$ at length $L$ as
\begin{equation}
    \Com(\A) = \{a\in A| a\prec b\cdot \bar{b} \text{ for some }b\in A\}.
\end{equation}
and inductively  
\begin{equation}
      \Com(\A)^L = \{a\in A| a\prec b_1\cdot \ldots \cdot b_L \text{ for some }b_i\in \Com(\A)\}
\end{equation}
Note that if $a\in \Com(\A)^L$, then $\overline{a}\in \Com(\A)^L$ and both $a,\overline{a}\in\Com(\A)^{L+1}$.

\paragraph{Loop level selection rules:} With these definitions, we can now state the loop level selection rules. A vertex $\cO=\phi_1 \phi_2 \cdots \phi_n$ is allowed at loop level $L$ if there exists some $\bar{b}\in \Com(\A)^L$ such that 
\begin{equation}
    e\prec a_1 \cdot a_2 \cdot\ldots\cdot a_n \cdot \bar{b}
\end{equation}
This is equivalent to the condition,
\begin{equation}
    \exists b\in\Com(\A)^L\text{ s.t.\ } b \pr a_1\cdot a_2\cdot\ldots\cdot a_n.
\end{equation}

For our purpose, we will only consider fusion algebras that admit new selection rules up to three loops. This is mathematically stated as $\mathrm{cl}(\A)=3$ where the commutator length $\mathrm{cl}(\A)$ is defined as the smallest integer $L$ such that $\Com(\A)^L=\Com(\A)^{L+1}$. With the restrictions we assume, namely that $N_{ab}^c\in \{0,1\}$, ref.~\cite{Kaidi:2024wio} compiled all possible independent fusion algebras up to rank 7. We first scan over these algebras and then turn to the longer list of fusion algebras up to rank 10 compiled at \cite{anyonwiki2024} (see \cite{Delaney:2026zwp} for an introduction). For $\mathrm{cl}(\A)\leq 3$ any term can first appear at tree, one-loop, two-loop or three-loop level, there are no new terms beyond that. Based on the fusion algebra alone, i.e. without a specific UV completion, one cannot estimate the numerical factors that modify the naive loop expansion. However, we implicitly assume the existence of a UV completion that admits a fusion algebra symmetry which, when restricted to the IR field content, reduces to the selection rules of the low energy theory.

\section{Yukawa textures from Fusion Algebras}
\label{sec:yukawa-setup}
We now outline how one can use a fusion algebra to generate Yukawa textures. 
% ============================================================
\subsection{Charged-Lepton Yukawa Textures}
\label{sec:lepton}
% ============================================================

First, consider the lepton sector of the Standard Model. The term that is responsible for the lepton masses is the Yukawa coupling term
\begin{equation}
    \Lag \supset Y^e_{ij} \bar{H} l_i e^c_j + \rm h.c.
\end{equation}
where $H$ is the Higgs doublet, $l_i,e^c_j$ are the left-handed lepton doublets and singlets respectively, and $Y^e_{ij}$ is the Yukawa matrix. We follow the two-component spinor notation as in ref.~\cite{Dreiner:2008tw}. The masses of the leptons are proportional to the singular values of the Yukawa matrix $Y^e$ where the proportionality constant is $|\langle H \rangle|$. The observed mass ratios of the charged leptons are $m_e:m_\mu:m_\tau \approx 1:210.7:3578$. Our goal is to see if we can reproduce this hierarchy from the fusion algebra selection rules. To do this, we assign to each field a basis element of the fusion algebra. We denote the assignments as 
\begin{equation}
    H^*[a],\quad l_i[b_i],\quad e^c_j[c_j]
\end{equation}
where $a,b_i,c_j\in A$ are basis elements of the fusion algebra. 

The entry $Y^e_{ij}$ is then determined by 
\begin{itemize}
    \item If $e\prec a b_i c_j$, then $Y^e_{ij}$ is allowed at tree level. We then assign $Y^e_{ij}\sim 1$.
    \item If it's not allowed at tree level but $\exists w\in\Com(\A)$ s.t. $w\prec a b_i c_j$, then $Y^e_{ij}$ is allowed at one-loop and we assign $Y^e_{ij}\sim \eps$.
    \item If it's not allowed at tree level or one-loop level  but $\exists w\in\Com(\A)^2$ s.t. $w\prec a b_i c_j$, then we assign $Y^e_{ij}\sim \eps^2$.
    \item If it's not allowed at tree level, one-loop or two-loop levels  but $\exists w\in\Com(\A)^3$ s.t. $w\prec a b_i c_j$, then we assign $Y^e_{ij}\sim \eps^3$.
\end{itemize}
Since all the algebras we consider are such that $\mathrm{cl}(\A)\leq 3$, there are no new selection rules beyond three loops. We then compute the singular values of the resulting Yukawa matrix. 

\paragraph{Benchmark value of $\eps$:} In order to compare the singular values to the observed mass ratios, we need to assign a value for $\eps$. For this, we choose the value 
\begin{equation}
    \eps = \sqrt{\frac{m_e}{m_\tau}} \approx 0.017
\end{equation}
In the remainder of the paper we quote every predicted mass ratio and mixing
element in two ways: as the decimal obtained by substituting the benchmark
$\eps=0.017$, and as its \emph{leading power} of $\eps$.  The decimals allow a
direct comparison with the measured values, while the powers of $\eps$ make
the dependence on the loop factor explicit.  

\subsection{An example calculation}
To illustrate the procedure, consider the algebra C.9 which has a commutator length $\rm{cl}(\A)=2$. The C.9 fusion ring on the basis $\{1,2,3,4,5,6\}$ is

\begin{center}
\renewcommand{\arraystretch}{1.2}
\begin{tabular}{c|cccccc}
$\times$ & 1 & 2 & 3 & 4 & 5 & 6\\\hline
1 & 1 & 2 & 3 & 4 & 5 & 6\\
2 & 2 & 3 & 1 & 5 & 6 & 4\\
3 & 3 & 1 & 2 & 6 & 4 & 5\\
4 & 4 & 5 & 6 & $3+4+6$ & $1+4+5$ & $2+5+6$\\
5 & 5 & 6 & 4 & $1+4+5$ & $2+5+6$ & $3+4+6$\\
6 & 6 & 4 & 5 & $2+5+6$ & $3+4+6$ & $1+4+5$\\
\end{tabular}
\end{center} 
Let us make the following assignment:
\begin{equation}
    H^*[1],\quad l_1[4],l_2[1],l_3[3].\quad e^c_1[6],e^c_2[2],e^c_3[3]
\end{equation}
Since $H$ is in the trivial representation, the only product that affects the loop level structure is of $l_i e^c_j$. The first step in identifying the loop order is to list the elements of $\Com (\rm C.9)$,
\begin{equation}
    \Com (\rm C.9) = \{ 1,4,5 \}
\end{equation}
We then list the different products and the order in which it appears:
\begin{center}
\renewcommand{\arraystretch}{1.35}
\begin{tabular}{|c| c| c| c|}
\hline
(i,j) & $l_i\cdot e^c_j$ & order & reason \\
\hline
(1,1) & $4\cdot 6 = 2+5+6$ & $\eps$   & $5\in\Com(\A)\Rightarrow$ 1-loop \\
(1,2) & $4\cdot 2 = 5$         & $\eps$   & $5\in\Com(\A)\Rightarrow$ 1-loop \\
(1,3) & $4\cdot 3 = 6$         & $\eps^2$ & $6\notin\Com(\A)\Rightarrow$ 2-loop \\
(2,1) & $1\cdot 6 = 6$         & $\eps^2$ & $6\notin\Com(\A)\Rightarrow$ 2-loop \\
(2,2) & $1\cdot 2 = 2 $        & $\eps^2$ & $2\notin\Com(\A)\Rightarrow$ 2-loop \\
(2,3) & $1\cdot 3 = 3 $        & $\eps^2$ & $3\notin\Com(\A)\Rightarrow$ 2-loop \\
(3,1) & $3\cdot 6 = 5 $        & $\eps$   & $5\in\Com(\A)\Rightarrow$ 1-loop \\
(3,2) & $3\cdot 2 = 1 $        & $1$      & Contains $1\Rightarrow$ tree \\
(3,3) & $3\cdot 3 = 2 $        & $\eps^2$ & $2\notin\Com(\A)\Rightarrow$ 2-loop \\
\hline
\end{tabular}
\end{center}
Therefore the resulting Yukawa matrix has the texture
\begin{equation}
    Y^e \sim \begin{pmatrix}
\eps & \eps & \eps^2\\
\eps^2 & \eps^2 & \eps^2\\
\eps & 1 & \eps^2
\end{pmatrix}
\end{equation}
The singular values of this matrix are in the ratio 
\begin{equation}
    m_e : m_\mu : m_\tau \approx 1 : 58.8 : 3522
    \qquad \left(\sim 1 : \eps^{-1} : \eps^{-2} \right)
\end{equation}
for the benchmark value of $\eps$ which provides an acceptable fit to the observed mass ratios, up to $\mathcal{O}(1)$ Wilson coefficient corrections. Note that there could be other assignments that reproduce this same hierarchy. 

\subsection{The quark sector}
\label{sec:quark-setup}
The lepton sector alone is not sufficient to test the applicability of this framework. Applying it in the quark sector is a stronger test for three reasons. First, there are two independent quark textures, for the up-type and down-type quarks. Second, the up quarks show a wider spread compared to the leptons and the down quarks. Third, quark mixing provides additional observables (in the form of the CKM matrix) that can be compared with predictions in addition to masses.
 The terms that are responsible for the quark masses are the Yukawa coupling terms
\begin{equation}
    \Lag \supset Y^u_{ij} H q_i u^c_j + Y^d_{ij} \bar{H} q_i d^c_j + \rm h.c.
\end{equation}
where $q_i$ are the left-handed quark doublets, $u^c_j,d^c_j$ are the right-handed up-type and down-type quark singlets respectively, and $Y^u_{ij},Y^d_{ij}$ are the up-type and down-type Yukawa matrices. The masses of the quarks are proportional to the singular values of the Yukawa matrices where the proportionality constant is $|\langle H \rangle|$. The observed mass ratios of the up-type quarks are $m_u:m_c:m_t \approx 1:506:137400$ and the observed mass ratios of the down-type quarks are $m_d:m_s:m_b \approx 1:19.9:1058$. 

\subsection{The neutrino sector}
\label{sec:weinberg-setup}

Finally, one can apply the fusion algebra approach to generate neutrino mass hierarchies. We assume that the neutrino masses are generated by the Weinberg operator
\begin{equation}
    \cO_W^{(ij)} \sim \frac{Y_\nu^{ij}}{\Lambda} (l_i H)(l_j H).
\end{equation}
 The neutrino sector is worth including since it is fixed by the lepton and Higgs labels that we have already chosen for the lepton sector. On the other hand, this is one place where the framework is applied to a higher dimensional operator. Given a representation for the leptons $l_i$ and the Higgs $H$, this would provide a selection rule for the neutrino mass matrix. 
 The neutrino mass matrix is then given by $M_\nu^{ij} = \frac{v^2}{\Lambda} Y_\nu^{ij}$.

 The Weinberg operator is a four-point operator, built from two lepton doublets $l_i[b_i]$ and two Higgs doublets $H[\bar{a}]$. Note that the operator contains $H$ and not $H^*$. Therefore, since we assigned $[a]$ to $H^*$, we assign $[\bar{a}]$ to $H$. The selection rule for the Weinberg operator is then given by the four-fold product $\bar{a}\cdot \bar{a}\cdot b_i\cdot b_j$. Hence, the operator is generated at loop level $L$ if there exists some $w\in\Com(\A)^L$ such that $w\prec \bar{a}\cdot \bar{a}\cdot b_i\cdot b_j$.
 
\subsection{The reach of a texture}
\label{sec:reach}

The loop-order dictionary has an immediate consequence.  Let $Y$ be a $3\times3$
matrix with entries $Y_{ij}=c_{ij}\eps^{n_{ij}}$ and
$0\leq n_{ij}\leq\mathrm{cl}(\A)$.  For generic $\cO(1)$ coefficients
$c_{ij}$, the largest singular value is $\cO(1)$ at best and the smallest is
$\cO(\eps^{\mathrm{cl}(\A)})$ at worst, so the attainable spread obeys
\begin{equation}
  \frac{m_3}{m_1}\lesssim\eps^{-\mathrm{cl}(\A)} .
  \label{eq:reach}
\end{equation}
We refer to $\eps^{-\mathrm{cl}(\A)}$ as the \emph{reach} of the algebra.  At
the benchmark $\eps=0.017$ the reach is $3.5\times10^{3}$ for
$\mathrm{cl}(\A)=2$ and $2.0\times10^{5}$ for $\mathrm{cl}(\A)=3$.  Comparing
with the observed spreads,
\begin{equation}
  \frac{m_\tau}{m_e}\approx3.6\times10^{3},\qquad
  \frac{m_b}{m_d}\approx1.1\times10^{3},\qquad
  \frac{m_t}{m_u}\approx1.4\times10^{5},
\end{equation}
we see that the charged-lepton and down-type sectors sit comfortably inside the
$\mathrm{cl}(\A)=2$ reach, while the up-type sector does not: it needs
$\eps^{-3}$.  Equation~\eqref{eq:reach} can of course be violated at
non-generic coefficients, and Sec.~\ref{sec:cl2} shows that this is exactly
what happens for C.9 --- which is why the bound must be tested against an
ensemble of coefficients rather than at $c_{ij}=1$.

\subsection{Scoring metrics}
\label{sec:metrics}

To compare algebras and assignments we use the following measures throughout. For a sector whose predicted singular values are in the ratio $1:\alpha:\beta$
and whose observed ratios are $1:\alpha_{\rm obs}:\beta_{\rm obs}$, we define
\begin{equation}
    \rho = \abs{\log_{10}\frac{\alpha}{\alpha_{\rm obs}}}
         + \abs{\log_{10}\frac{\beta}{\beta_{\rm obs}}} ,
    \label{eq:rho}
\end{equation}
so that $\rho=0$ is a perfect fit $\rho$ increasing by one corresponds to the predicted ratios moving away from the observed ones by one order of magnitude (in total).  We write $\rho_\ell$, $\rho_u$, $\rho_d$ for
the charged-lepton, up-type and down-type scores.  For the mixing matrix we use
\begin{equation}
    \rho_{\rm CKM} = \frac{1}{9}\sum_{i,j}
    \abs{\log_{10}\frac{|V^{\rm alg}_{ij}|}{|V^{\rm obs}_{ij}|}} ,
    \label{eq:rho-ckm}
\end{equation}
and we
combine the two quark-mass scores into
\begin{equation}
    \rho_{\rm mass} = \sqrt{\frac{\rho_{\rm up}^2 + \rho_{\rm down}^2}{2}} .
    \label{eq:rho-mass}
\end{equation}
Finally, the combined figure of merit used to select a single assignment across
all sectors is
\begin{equation}
  \rho_{\rm tot} =
  \sqrt{\tfrac{1}{4}\bigl(\rho_\ell^{2}+\rho_u^{2}+\rho_d^{2}
        +\rho_{\rm CKM}^{2}\bigr)} .
  \label{eq:rho-tot-combined}
\end{equation}

\subsection{Beyond unit Wilson coefficients}
\label{sec:ensemble}

Every ratio quoted from a texture alone assumes coefficients of the form
$1\times\eps^{n}$.  To show that a hierarchy is a genuine consequence of the
algebra and not a coincidence of a fine-tuned $\cO(1)$ coefficient, we test
each texture against an ensemble.  We scan the ranges

\begin{equation}
  Y_{ij} = s_{ij}|c_{ij}|\eps^{n_{ij}},
  \qquad
  \log|c_{ij}| \sim U\bigl[-\log 2.5, +\log 2.5\bigr],
  \qquad
  s_{ij}=\pm 1 
  \label{eq:ensemble}
\end{equation}
and quote the median and the $68\%$ band of every observable over the
ensemble.  The difference between the median and the value at $c_{ij}=1$ is
precisely the diagnostic for an accidental cancellation, and in
Sec.~\ref{sec:cl2-failure} it is what disqualifies the $\mathrm{cl}(\A)=2$
up-type fit. Note that, for the observed values, the uncertainties are small compared to the scores and widths of the 68\% bands over the coefficient ensemble. We therefore do not propagate the experimental uncertainties into the fit.

% ============================================================
\section{$\rm{cl}(\A)=2$ Selection Rules and Their Limits}
\label{sec:cl2}
% ============================================================

Every multiplicity-free fusion algebra of rank $\leq 7$ in the classification of
ref.~\cite{Kaidi:2024wio} has $\mathrm{cl}(\A)\leq 2$.  This section summarises
what such algebras achieve and where they fail.  The detailed analysis --- the
worked C.9 example, the per-sector scans over all nineteen algebras, the
neutrino spectra and the CKM fits --- is collected in Appendix~\ref{app:cl2},
and the fusion tables in Appendix~\ref{app:classification}.

\subsection{What a $\rm{cl}(\A)=2$ rule achieves}
\label{sec:cl2-results}

Scanning all assignments of $(H,\ell_i,e^c_j,q_i,u^c_j,d^c_j)$ in all nineteen
algebras and scoring with Eq.~\eqref{eq:rho}, the outcome is easy to state.

\begin{itemize}
  \item \textbf{A two-stage rule produces a two-stage hierarchy.}  Entries in
        $\{1,\eps,\eps^2\}$ generate singular values in the pattern
        $1:\eps^{-1}:\eps^{-2}$, which is the qualitative shape of the observed
        charged-lepton and down-type spectra.  This is the basic success of the
        framework and it does not depend on which algebra is chosen.
  \item \textbf{One algebra stands out: the rank-6, non-categorifiable algebra
        C.9.}  With $H^*[1]$, $\ell_i=(4,1,3)$ and $e^c_j=(6,2,3)$ it gives
        $m_e:m_\mu:m_\tau\approx1:58.8:3522$ against the observed
        $1:210.7:3578$ ($\rho_\ell=0.561$), and with $q_i=\ell_i$,
        $d^c_j=e^c_j$ the same texture gives $m_d:m_s:m_b\approx1:58.8:3522$
        against $1:19.9:1058$ ($\rho_d=0.994$).  It is also the only
        rank-$\leq7$ algebra whose up-type sector reaches $\eps^{-3}$ at unit
        coefficients - this is shown in detail below. The full comparison is in
        Appendix~\ref{app:cl2-lepton-scan} and Sec.~\ref{sec:quark}.
  \item \textbf{The neutrino sector prefers the same algebra.}  Through the
        Weinberg operator of Sec.~\ref{sec:weinberg-setup}, C.9 gives a normal
        hierarchy $1:\eps:\eps^2$ with
        $\Delta m^2_{31}\approx2.5\times10^{-3}\mathrm{eV}^2$ at
        $\Lambda\approx6\times10^{14}$~GeV, while the algebras that fit the
        charged leptons through a degenerate $1:1:1$ spectrum leave
        $\Delta m^2_{21}$ unsuppressed (Appendix~\ref{app:cl2-neutrino}).
  \item \textbf{There is no good fit to the CKM matrix.}  The
        mass-optimal C.9 assignment gives a near-maximal $1$--$2$ rotation,
        $|V_{us}|\sim 0.70$ against the observed $0.225$, and no $\cO(1)$
        coefficients of natural size repair it. On the other hand, the CKM-optimal assignment
        reaches a median $\rho_{\rm CKM}=0.316$ but degrades  the median $\rho_{\rm mass}$ from
        $1.79$ to $2.19$ (Sec.~\ref{sec:CKM}).
\end{itemize}

\subsection{Why the up-type sector fails}
\label{sec:cl2-failure}

The C.9 up-type fit with unit Wilson coefficients,
$m_u:m_c:m_t\approx1:7162:2.98\times10^{5}$, exceeds the reach bound
\eqref{eq:reach} for $\mathrm{cl}(\A)=2$.  It therefore has to be an accident,
and it is.
Let us express the up type Yukawa matrix as $Y^u_{ij}\equiv c_{ij}\eps^{n_{ij}}$, then the texture that gives the best hierarchy among up type quarks is the following:
\begin{equation}
    Y^u \sim \begin{pmatrix}
      c_{11}\eps   & c_{12}\eps^2 & c_{13}\eps\\
      c_{21}\eps   & c_{22}\eps^2 & c_{23}\eps^2\\
      c_{31}\eps^2 & c_{32}\eps^2 & c_{33}\eps^0
    \end{pmatrix}
\end{equation}
We see that 
\begin{equation}
    \det Y^u = c_{33}(c_{11}c_{22}-c_{12}c_{21})\eps^3+\cO(\eps^4).
\end{equation}
At $c_{ij}=1$ the bracket vanishes identically and gives us an $\cO(\eps^4)$ determinant. This accidental cancellation is what is providing us with a broader texture.

The results for the C.9 texture are summarized in Table~\ref{tab:mc_c9}.
\begin{table}[H]
\centering\small
\renewcommand{\arraystretch}{1.2}
\begin{tabular}{|l|c|c|c|c|c|}
\hline
\multirow{2}{*}{Sector} & \multicolumn{3}{c|}{$m_3/m_1$} &
\multicolumn{2}{c|}{$\rho$}  \\
\cline{2-6}
 & observed & at $c=1$ & MC median & at $c=1$ & MC median \\
\hline
C.9 charged lepton & $3578$            & $3522$             & $3847$ & $0.561$ & $0.722$  \\
C.9 down-type      & $1058$            & $3522$             & $3460$ & $0.994$ & $0.988$  \\
C.9 up-type & $1.37\times10^{5}$ & $2.98\times10^{5}$ & $3854$ & $1.487$ & $2.290$ \\
\hline
\end{tabular}
\caption{The C.9 assignment of Secs.~\ref{sec:lepton} and \ref{sec:quark} under
the ensemble of Eq.~\eqref{eq:ensemble}. Note that the median among the MC scan is $\cO(\eps)$ lower than the fine-tuned result.}
\label{tab:mc_c9}
\end{table}

The charged-lepton and down-type fits are therefore genuine predictions of the
selection rule. The up-type fit on the other hand is
not: its median $m_t/m_u$ is $3.85\times10^{3}$ against the observed
$1.37\times10^{5}$. This is the expected consequence of $\rm{cl}(\A)=2$. Since the entries are drawn from $\{1,\eps,\eps^{2}\}$ a generic matrix cannot spread its
singular values by more than $\eps^{-\rm{cl}(\A)}$, while
$m_t/m_u\approx1.37\times10^{5}$ needs $\eps^{-3}$ at $\eps=0.017$. The
top-quark hierarchy is exactly what a two-stage rule cannot accommodate. 

This is the obstruction that organises the rest of the paper.  It is not a
statement about C.9 in particular, nor about rank: it is the reach bound
\eqref{eq:reach} applied to $\mathrm{cl}(\A)=2$.  No algebra with a two-stage
selection rule can generate the up-quark hierarchy for generic $\cO(1)$
coefficients, and every algebra of rank $\leq7$ has a two-stage selection rule.

\section{The unique algebra of rank$\leq 9$ with $\rm{cl}(\A)=3$}
\label{sec:rank8}
In order to explain the up-quark hierarchy, we need a $\rm{cl}(\A)=3$ fusion algebra. For algebras with rank less than or equal to $7$, the maximum value of $\rm{cl}(\A)$ is $2$. Hence, we need to consider higher rank algebras. For this, we considered the fusion ring catalogue given by Vercleyen and Slingerland~\cite{vercleyen2023lowrankfusionrings}. Computing the commutator length for all fusion algebras of ranks up to $9$, we find that there is a unique algebra\footnote{ Carrying the same analysis one step further, to rank $10$, turns up three more algebras with $\rm{cl}(\A)=3$ and still none with $\rm{cl}(\A)>3$. They are analysed in Appendix~\ref{app:rank10}, and the upshot is that they do not improve on $\mathrm{FR}^{8,6}_{8}$ in any sector except, marginally, the CKM matrix.} with $\rm{cl}(\A)=3$ and none with $\rm{cl}(\A)>3$. The algebra is of rank $8$, is non-categorifiable and is the ring $\mathrm{FR}^{8,6}_{8}$. The fusion table is 
\begin{center}
\scriptsize
\renewcommand{\arraystretch}{1.15}
\begin{tabular}{c|cccccccc}
$\times$ & $1$ & $2$ & $3$ & $4$ & $5$ & $6$ & $7$ & $8$ \\
\hline
$1$ & $1$ & $2$ & $3$ & $4$ & $5$ & $6$ & $7$ & $8$ \\
$2$ & $2$ & $1$ & $4$ & $3$ & $8$ & $7$ & $6$ & $5$ \\
$3$ & $3$ & $4$ & $2$ & $1$ & $6$ & $8$ & $5$ & $7$ \\
$4$ & $4$ & $3$ & $1$ & $2$ & $7$ & $5$ & $8$ & $6$ \\
$5$ & $5$ & $8$ & $6$ & $7$ & $4{+}6{+}8$ & $1{+}7{+}8$ & $2{+}5{+}6$ & $3{+}5{+}7$ \\
$6$ & $6$ & $7$ & $8$ & $5$ & $1{+}7{+}8$ & $3{+}5{+}7$ & $4{+}6{+}8$ & $2{+}5{+}6$ \\
$7$ & $7$ & $6$ & $5$ & $8$ & $2{+}5{+}6$ & $4{+}6{+}8$ & $3{+}5{+}7$ & $1{+}7{+}8$ \\
$8$ & $8$ & $5$ & $7$ & $6$ & $3{+}5{+}7$ & $2{+}5{+}6$ & $1{+}7{+}8$ & $4{+}6{+}8$
\end{tabular}
\end{center}
Its invertible objects
$\{1,2,3,4\}$ form a $\mathbb{Z}_4$, the duals are $\bar3=4$, $\bar5=6$,
$\bar7=8$ with $1,2$ self-dual, and the filtration is
\begin{equation}
\Com(\A)=\{1,7,8\}
\ \subsetneq\
\Com(\A)^2=\{1,3,4,5,6,7,8\}
\ \subsetneq\
\Com(\A)^3=A,
\qquad \rm{cl}(\A)=3.
\label{eq:rank8-filtration}
\end{equation}

\subsection{The best assignment}
\label{sec:rank8-best}

We can now ask for whether there is an assignment that fits all the textures at once. For this, we define the combined metric,
\begin{equation}
  \rho_{\rm tot} \;=\;
  \sqrt{\tfrac{1}{4}\bigl(\rho_\ell^{2}+\rho_u^{2}+\rho_d^{2}
        +\rho_{\rm CKM}^{2}\bigr)} ,
  \label{eq:rho-tot}
\end{equation}
Then, within one
draw of assignments, we build all three Yukawa matrices with a Monte-Carlo scan of the $\cO(1)$ coefficients,
score each sector against the observed ratios with the metric of
Eq.~\eqref{eq:rho}, form $V_{\rm CKM}$ from $Y^u$ and $Y^d$
minimising the median of $\rho_{\rm tot}$ over the ensemble, not its
value at $c_{ij}=1$. The outcome for $\mathrm{FR}^{8,6}_{8}$ is
\begin{equation}
  H^{*}[4],\qquad
  \ell_i=(2,4,8),\quad e^c_j=(2,3,7),\qquad
  q_i=(2,4,6),\quad u^c_j=(4,7,8),\quad d^c_j=(1,5,6),
  \label{eq:rank8-best}
\end{equation}
giving the textures
\begin{equation}
Y^{e}\sim
\begin{pmatrix}
\eps^{2}&\eps^{3}&\eps^{2}\\
1&\eps^{2}&\eps^{2}\\
\eps&\eps&\eps\end{pmatrix},
\qquad
Y^{u}\sim
\begin{pmatrix}\eps^{3}&\eps&\eps^{2}\\
\eps^{2}&\eps&\eps\\
\eps^{2}&1&\eps
\end{pmatrix},
\qquad
Y^{d}\sim
\begin{pmatrix}
\eps^{2}&\eps^{2}&\eps\\
\eps^{3}&\eps&\eps\\
\eps^{2}&\eps&1
\end{pmatrix},
\label{eq:rank8-textures}
\end{equation}
and the scores of Table~\ref{tab:rank8-scores}.

\begin{table}[H]
\centering\small
\renewcommand{\arraystretch}{1.2}
\begin{tabular}{|l|c|c||l|c|c|}
\hline
Score & median & $68\%$ band & Spectrum & MC median & observed \\
\hline
$\rho_\ell$        & $0.601$ & $[0.306,0.950]$ & $m_e:m_\mu:m_\tau$ & $1:146:5416$ & $1:210.7:3578$ \\
$\rho_u$           &$1.000$ & $[0.574,1.699]$ & $m_u:m_c:m_t$ & $1:1773:1.05{\times}10^{5}$ & $1:506:1.37{\times}10^{5}$ \\
$\rho_d$           & $0.986$ & $[0.457,\,1.587]$ & $m_d:m_s:m_b$ & $1:59.0:3464$ & $1:19.9:1058$ \\
$\rho_{\rm CKM}$   & $0.408$ & $[0.288,0.576]$ & \multicolumn{3}{c|}{} \\
\hline
$\boldsymbol{\rho_{\rm tot}}$ & $\mathbf{0.885}$ & $[0.644,1.196]$
& \multicolumn{3}{c|}{} \\
\hline
\end{tabular}
\caption{The assignment of Eq.~\eqref{eq:rank8-best} under the ensemble of
Eq.~\eqref{eq:ensemble}. }
\label{tab:rank8-scores}
\end{table}

The generic reaches of the three textures of Eq.~\eqref{eq:rank8-textures} are
$\eps^{-2}$, $\eps^{-3}$ and $\eps^{-2}$ for $Y^e$, $Y^u$ and $Y^d$
respectively, and in each case the Monte-Carlo median of Table
\ref{tab:rank8-scores} agrees with what that reach predicts. Therefore, nothing here is
carried by an accidental cancellation. The top-quark hierarchy comes out at
$1.05\times10^{5}$ against the observed $1.37\times10^{5}$ because the selection
rule allows $\eps^{-3}$ and the assignment uses it.

\subsection{The CKM matrix for $\mathrm{FR}^{8,6}_{8}$}
\label{sec:rank8-ckm}
In this section, we will discuss the CKM structure that arises from the Yukawa matrices with the textures we get from the fusion algebra structure. The CKM matrix is given by $V_{\rm CKM} =  V^{u\dagger}_L V^d_L$ where $V^u_L$ and $V^d_L$ are the unitary matrices that diagonalize the up-type and down-type Yukawa matrices respectively by a singular value decomposition,
\begin{equation}
    Y^u=V^u_L \Sigma^u V^{u\dagger}_R,\quad Y^d=V^d_L \Sigma^d V^{d\dagger}_R
\end{equation}
with $\Sigma^u$ and $\Sigma^d$ being the diagonal matrices with the singular values of $Y^u$ and $Y^d$ respectively. We order the columns of $V^u_L$ and $V^d_L$ such that the singular values are in increasing order so that the first, second and third generations correspond to $u,c,t$ and $d,s,b$ respectively.

Diagonalising the two quark textures of Eq.~\eqref{eq:rank8-textures} across
the ensemble gives
\begin{equation}
    |V_{\rm CKM}| =
    \begin{pmatrix}
        0.9985^{+0.0012}_{-0.0046} & 0.040^{+0.058}_{-0.029} & 0.026^{+0.029}_{-0.019} \\
        0.040^{+0.058}_{-0.029} & 0.9984^{+0.0013}_{-0.0044} & 0.026^{+0.029}_{-0.019} \\
        0.026^{+0.029}_{-0.019} & 0.026^{+0.029}_{-0.019} & 0.9991^{+0.0007}_{-0.0020}
    \end{pmatrix},
    \qquad \rho_{\rm CKM} = 0.408 ,
    \label{eq:ckm-rank8}
\end{equation}
where the error bars correspond to the $68\%$ band as shown in Table~\ref{tab:rank8-scores}.

To compare with, we also give the observed CKM matrix, which is given by~\cite{ParticleDataGroup:2024cfk} 
\begin{equation}
 |V_{\rm CKM}^{\rm obs}| \approx
 \begin{pmatrix}
 0.9744 & 0.2250 & 0.0037\\
 0.2249 & 0.9735 & 0.0418\\
 0.0086 & 0.0411 & 0.9991
 \end{pmatrix},
 \label{eq:ckm-obs}
\end{equation}
We see that some of the qualitative features of the CKM matrix are captured by the fusion algebra structure, such as the fact that the third generation is mostly decoupled from the first two generations.Comparing with Eq.~\eqref{eq:ckm-obs}, the near-unit diagonal and the
third-generation mixings are reproduced at the few-percent level.  The
$1$--$2$ mixing remains the weak point: the ensemble gives
$|V_{us}|=0.040^{+0.058}_{-0.029}$ against the observed $0.225$, so the
near-maximal Cabibbo angle of the $\mathrm{cl}(\A)=2$ textures
(Sec.~\ref{sec:CKM}) has been traded for one that is too small.

 It would be interesting to ask if there are directions on the coefficient space that takes us closer to the observed CKM matrix. To see, we look at the top $1\%$ of draws, which gives a median $\rho_{\rm CKM}=0.134$, with the best draw giving $\rho_{\rm CKM} = 0.026$. Looking at the top draws, it is not clear if there are specific directions in the coefficient space that makes a better fit. In other words, there is no appreciable correlation between $\rho_{\rm CKM}$ and any individual entry in the coefficient space. But it is very clear that the observed CKM matrix is almost perfectly achieved within the draws, with a score of $\rho_{\rm CKM} = 0.026$. Whether this choice is correlated with other observables, such as the quark mass ratios, could be addressed with a multivariate analysis of the ensemble, which we leave for future work. 

\section{Neutrinos from $\mathrm{FR}^{8,6}_{8}$}
\label{sec:rank8-neutrino}

 We now look at what this assignment predicts for the neutrino sector.
The Weinberg-operator selection rule of Sec.~\ref{sec:weinberg-setup} applies
unchanged: with $\ell_i[b_i]$ and $H[\bar a]$, the operator $\cO_W^{(ij)}$ is
generated at loop order $L$ if some $w\in\Com(\A)^L$ satisfies
$w\prec\bar a\,\bar a\,b_i\,b_j$.
 
  Unlike every other sector, the
neutrino sector involves no choice: the assignment of
Eq.~\eqref{eq:rank8-best} fixes $H^{*}[4]$ and $\ell_i=(2,4,8)$, and the
Weinberg operator introduces no new fields. 

\paragraph{The predicted texture:} 
Since $\overline{4}=3$ in $\mathrm{FR}^{8,6}_{8}$, the Higgs in
$\cO_W$ carries the label $3$, and the relevant product is
$3\cdot3\cdot b_i\cdot b_j = 2\cdot b_i\cdot b_j$.  Evaluating it on
$\ell_i=(2,4,8)$ gives
\begin{equation}
    M_\nu \sim \frac{v^2}{\Lambda}
    \begin{pmatrix}
      \eps^{3} & \eps^{2} & \eps\\
      \eps^{2} & 1        & \eps\\
      \eps     & \eps     & \eps
    \end{pmatrix},
    \label{eq:nu-texture-rank8}
\end{equation}
symmetric, as the commutativity of the fusion algebra guarantees. Since this texture predicts a split between the highest eigenvalues, inverted ordering for the neutrinos is disfavored. Henceforth, we will assume normal ordering.

The Weinberg operator is often taken to be generated by integrating out some heavy SM singlet fermions, $N_a$, with 
\begin{equation}
\mathcal{L} = \dots - y^{ia} (l_i H)N_a - \Lambda \gamma^{ab}N_a N_b + \dots
\end{equation}
so that
\begin{equation}
Y^{ij}_\nu = y^{ia}y^{jb} (\gamma^{-1})_{ab}
\end{equation}
In the present context, we can for example assign the labels $N_1[1]$, $N_2[7]$ and $N_3[8]$. This produces a Majorana mass matrix with texture
\begin{equation}
\gamma^{ab}\sim\begin{pmatrix}
1&\eps &\eps\\
\eps &\eps &1\\
\eps&1&\eps
\end{pmatrix}
\end{equation}
and Yukawa matrix with texture
\begin{equation}
y^{ia}\sim\begin{pmatrix}
\eps^2 &\eps &\eps^2\\
1 &\eps &\eps \\
\eps &\eps & 1
\end{pmatrix}
\end{equation}
Putting these together, we reproduce
\begin{equation}
\begin{split}
M_\nu^{ij}&= \frac{v^2}{\Lambda}(y\cdot \gamma^{-1}\cdot y^t)^{ij}\\
&\sim  \frac{v^2}{\Lambda}
    \begin{pmatrix}
      \eps^{3} & \eps^{2} & \eps\\
      \eps^{2} & 1        & \eps\\
      \eps     & \eps     & \eps
    \end{pmatrix},
\end{split}
\end{equation}
 
Table~\ref{tab:rank8-nu} gives the spectrum under the ensemble of
Eq.~\eqref{eq:ensemble}, with one modification required by the physics: the
coefficients are drawn symmetrically, $c_{ij}=c_{ji}$, since $M_\nu$ is a
Majorana mass matrix and nine independent coefficients would sample matrices no
Lagrangian produces.
 
\begin{table}[H]
    \centering
    \small
    \renewcommand{\arraystretch}{1.2}
    \begin{tabular}{|l|c|c|c|c|}
        \hline
        Quantity & at $c_{ij}=1$ & MC median & $68\%$ band & observed \\
        \hline
        $m_{\nu_2}/m_{\nu_3}$ & $0.0273$ & $0.0301$ &
        $[0.0152,\,0.0610]$ & --- \\

        $m_{\nu_1}/m_{\nu_3}$ & $0.0106$ & $0.0101$ &
        $[0.0034,\,0.0266]$ & --- \\

        $\Delta m^{2}_{21}/\Delta m^{2}_{31}$ & $6.3\times10^{-4}$ &
        $7.0\times10^{-4}$ & $[1.7,\,28.5]\times10^{-4}$ &
        $3.0\times10^{-2}$ \\

        $\Lambda$ [GeV] & $6.1\times10^{14}$ & $6.1\times10^{14}$ &
        $[3.3,\,11.3]\times10^{14}$ & --- \\
        \hline
    \end{tabular}
    \caption{The neutrino sector predicted by the assignment of
    Eq.~\eqref{eq:rank8-best}, from $4\times10^{5}$ draws of the ensemble of
    Eq.~\eqref{eq:ensemble} with symmetric coefficients. The scale $\Lambda$ is
    fixed by $m_{\nu_3}=0.05$~eV and carries no information; the observable that
    tests the algebra is the splitting ratio. Normal ordering is assumed
    throughout. Observed value from the NuFIT~6.0 global fit
    \cite{Esteban:2024eli}.}
    \label{tab:rank8-nu}
\end{table}
 
What we find is $\Delta m^{2}_{21}/\Delta m^{2}_{31}=7.0\times10^{-4}$,
a factor of $43$ below the observed $3.0\times10^{-2}$.  The corresponding C.9
number is $2.9\times10^{-4}$, low by a factor of $105$, so rank 8 improves
matters by a factor of about two and a half. The scale comes out
at $\Lambda\approx6\times10^{14}$~GeV, the usual GUT-scale statement, but this is
fixed by hand and is not a test.

\subsection{The PMNS matrix}
Diagonalising $Y^e$ of Eq.~\eqref{eq:rank8-textures} and $M_\nu$ of
Eq.~\eqref{eq:nu-texture-rank8} across the ensemble, with independent
coefficients for the two operators, gives
\begin{equation}
    |U_{\rm PMNS}| =
    \begin{pmatrix}
        0.851^{+0.082}_{-0.070} & 0.525^{+0.100}_{-0.166} & 0.001^{+0.001}_{-0.000} \\[2pt]
        0.525^{+0.099}_{-0.166} & 0.850^{+0.082}_{-0.070} & 0.026^{+0.029}_{-0.019} \\[2pt]
        0.012^{+0.016}_{-0.009} & 0.022^{+0.025}_{-0.016} & 1.000^{+0.000}_{-0.001}
    \end{pmatrix},
    \label{eq:pmns-rank8}
\end{equation}
quoting the median with its $68\%$ band, against the observed magnitudes
\cite{Esteban:2024eli}
\begin{equation}
|U^{\rm obs}_{\rm PMNS}| \approx
\begin{pmatrix}
0.823 & 0.548 & 0.148\\
0.275 & 0.613 & 0.741\\
0.497 & 0.569 & 0.655
\end{pmatrix}.
\label{eq:pmns-obs}
\end{equation}
 
We see that $|U_{e1}|=0.851$ and
$|U_{e2}|=0.5$ are reproduced well against the measured $0.823$ and $0.548$, with nothing fitted. The rest is not: $|U_{e3}|$ is smaller than observed by two orders of
magnitude, and the $2$--$3$ block comes out almost unmixed where the data
require near-maximal mixing.  The predicted matrix is close to block diagonal
in the $2$--$3$ sector. 

 Unlike the case of CKM matrix, here the MC scan does identify a direction that takes us closer to the observed magnitudes. Again, looking at the top $1\%$ of draws, we find that $c_{22}$ is pulled down while $c_{23}$ is pulled up. This makes sense since we expect such a pull to improve the mixing between the second and third generation. The score
improves from a median of $0.857$ to $0.532$ 

% ============================================================
\section{Discussion}
\label{sec:discussion}
% ============================================================

The fusion-algebra framework explains the fermion mass hierarchy in a way that differs
structurally from the familiar flavour mechanisms. In Froggatt--Nielsen models~\cite{Froggatt:1978nt,Leurer:1992wg,Leurer:1993gy} and in
non-Abelian or modular flavour models~\cite{Altarelli:2010gt,Ishimori:2010au,Feruglio:2017spp},
the small parameter is the vacuum expectation value of a new field, and the charges or
representations are chosen to place its powers in the right entries. Here, we find that the suppression is equally well applicable as possibly from a flavon or as a conseqence of UV loop effects. The hierarchy is fixed entirely by the multiplication table of $\A$ and by which
simple object each field carries. A single parameter, the ``loop'' factor $\eps$, is common to all
three charged-fermion sectors and to the neutrino sector. Every entry of every Yukawa matrix is
then $\eps^{n}$ with an integer $n\in\{0,\dots,\mathrm{cl}(\A)\}$ that the algebra computes rather
than one that is chosen. The textures also have no zeros since $\Com(\A)^{\mathrm{cl}(\A)}=A$
for the algebras we use, every operator is generated at some finite loop order. The hierarchy is
therefore graded rather than enforced by vanishing entries. We also show how the best fit textures remain stable under RGE.

The commutator length $\mathrm{cl}(\A)$ is the single property of the algebra that decides which
spectra are reachable. A texture with exponents in $\{0,\dots,\mathrm{cl}(\A)\}$ cannot spread its
singular values by more than $\eps^{-\mathrm{cl}(\A)}$ for generic $\cO(1)$ coefficients
(Eq.~\eqref{eq:reach}). This turns
the up-quark spectrum into a direct probe of the depth of the symmetry. With
$m_t/m_u\simeq1.4\times10^{5}$ and $\eps^{-2}\simeq3.5\times10^{3}$, a two-stage selection rule
is excluded for generic coefficients, and the top quark requires $\mathrm{cl}(\A)\ge3$. The
Monte-Carlo ensemble makes this quantitative. The best $\mathrm{cl}(\A)=2$ fit reproduces the
up-type hierarchy only through an exact cancellation at unit coefficients, and its ensemble
median collapses to $m_t/m_u\simeq3.9\times10^{3}$ (Sec.~\ref{sec:cl2-failure}).

Because the selection rule is determined by a fusion ring, the space of models is not a continuum
of charges but a finite list that can be searched exhaustively. Every multiplicity-free algebra of
rank $\le7$~\cite{Kaidi:2024wio} has $\mathrm{cl}(\A)\le2$. Among all multiplicity-free commutative
rings of rank $\le10$~\cite{vercleyen2023lowrankfusionrings,anyonwiki2024}, exactly four have
$\mathrm{cl}(\A)=3$ and none has $\mathrm{cl}(\A)>3$. 
Below rank 10 the requirement $\mathrm{cl}(\A)\ge3$ has a unique solution,
$\mathrm{FR}^{8,6}_{8}$. Within that one algebra, a single assignment of labels reproduces all
three charged-fermion hierarchies at the level of the generic reach. The three rank-10 algebras with
$\mathrm{cl}(\A)=3$ do not improve the mass fits (Appendix~\ref{app:rank10}). 

Once the lepton doublet and Higgs labels are fixed by the charged-lepton fit, the Weinberg
operator's texture is fully determined. For $\mathrm{FR}^{8,6}_{8}$ it predicts a normal hierarchy with a scale $\Lambda\sim6\times10^{14}$~GeV, 
 $|U_{e1}|\simeq0.85$ and $|U_{e2}|\simeq0.53$, against the measured $0.82$ and $0.55$, with
  nothing fitted (Sec.~\ref{sec:rank8-neutrino}).
The failures are equally sharp. $\Delta m^2_{21}/\Delta m^2_{31}$ is too small by a factor of about
40, and the $2$--$3$ block is almost unmixed. 

A fusion-algebra texture is generated at the scale $\Lambda$ of the heavy states. In principle, it could be spoiled by RGE running down to lower energies. Below $\Lambda$ only Standard Model loops remain. Every flavour-changing term in the one-loop Yukawa $\beta$-functions is a chain
$Y^\beta Y^{\beta\dagger}Y^\alpha$ closed by a light Higgs line. If the Higgs carries an invertible simple object, $h\bar h=1$, we have shown in Appendix~\ref{app:RGstab} that the texture is stable under 1-loop RGE running and survives from the scale $\Lambda$ down to the electroweak scale. This is a special feature of the textures generated by a fusion algebra symmetry.

Several questions remain open at this point. $\mathrm{FR}^{8,6}_{8}$ is not the Grothendieck ring of any fusion category. The usual realisations
of non-invertible symmetries, through topological defects, TQFTs or string worldsheet orbifolds, are
therefore not available for it. For the three rank-10 algebras with $\mathrm{cl}(\A)=3$,
categorifiability is, to our knowledge, open. It would be interesting to explore the UV completions of non-categorifiable fusion algebras.

This is particularly pressing if one wants to go beyond the crude analysis of this paper and fix the $O(1)$ numbers which multiply each power of $\eps$ in our textures. In particular, these $O(1)$ numbers are in general \emph{complex}, and their phases are crucial if we want to extend the search to include the CKM and PMNS phases. We leave these obvious extensions to future work.

\appendix

\section{Quick Reference: All Fusion Algebra Tables}
\label{app:classification}
% ===========================================================================

This Appendix tabulates the fusion rules for the commutative algebras\footnote{We only consider commutative fusion algebras; the $C.4$ algebra in \cite{Kaidi:2024wio} is non-commutative, so we omit it.},  with $\mathrm{cl}(\mathcal{A}) = 2$,  from
Appendix~C of Kaidi--Tachikawa--Zhang \cite{Kaidi:2024wio}.
For each algebra we give its fusion table, whether it is categorifiable and list its Frobenius-Perron dimension (an algebraic number which is a rough measure of its complexity). We also provide its name in the conventions of \href{AnyonWiki.org}{AnyonWiki.org}. Finally, we provide the same data for $\mathrm{FR}^
{8,6}_{8}$, the unique rank $\leq 9$ multiplicity-free algebra with $\mathrm{cl}(\mathcal{A}) \geq 3$.

% ─── Rank-4 ──────────────────────────────────────────────────────────────────

\subsection*{Rank-4 Algebras}

\subsubsection*{C.1a (\href{https://anyonwiki.org/rings/4_1_2_4}{$\text{FR}^{4,1,2}_{4}$}) --- Categorifiable, $D_{\text{FP}}^2 =8+4\sqrt{2}$}

\begin{center}
\small
\begin{tabular}{c|cccc}
$\times$ & $1$ & $2$ & $3$ & $4$ \\
\hline
$1$ & $1$ & $2$ & $3$ & $4$ \\
$2$ & $2$ & $2{+}3{+}4$ & $1{+}2{+}3$ & $3$ \\
$3$ & $3$ & $1{+}2{+}3$ & $2{+}3{+}4$ & $2$ \\
$4$ & $4$ & $3$ & $2$ & $1$ \\
\end{tabular}
\end{center}

\subsubsection*{C.1b (\href{https://anyonwiki.org/rings/4_1_0_4}{$\text{FR}^{4,1,0}_{4}$}) --- Categorifiable (HI($\mathbb{Z}_2$)/PSU(2)$_6$), $D_{\text{FP}}^2 =8+4\sqrt{2}$}

\begin{center}
\small
\begin{tabular}{c|cccc}
$\times$ & $1$ & $2$ & $3$ & $4$ \\
\hline
$1$ & $1$ & $2$ & $3$ & $4$ \\
$2$ & $2$ & $1$ & $4$ & $3$ \\
$3$ & $3$ & $4$ & $1{+}3{+}4$ & $2{+}3{+}4$ \\
$4$ & $4$ & $3$ & $2{+}3{+}4$ & $1{+}3{+}4$ \\
\end{tabular}
\end{center}

% ─── Rank-5 ──────────────────────────────────────────────────────────────────

\subsection*{Rank-5 Algebra}

\subsubsection*{C.2 (\href{https://anyonwiki.org/rings/5_1_0_5}{$\text{FR}^{5,1,0}_{5}$})--- Non-categorifiable, $D_{\text{FP}}^2 =13+\sqrt{13}$}

\begin{center}
\small
\begin{tabular}{c|ccccc}
$\times$ & $1$ & $2$ & $3$ & $4$ & $5$ \\
\hline
$1$ & $1$ & $2$ & $3$ & $4$ & $5$ \\
$2$ & $2$ & $1$ & $5$ & $4$ & $3$ \\
$3$ & $3$ & $5$ & $1{+}4{+}5$ & $3{+}5$ & $2{+}3{+}4$ \\
$4$ & $4$ & $4$ & $3{+}5$ & $1{+}2{+}4$ & $3{+}5$ \\
$5$ & $5$ & $3$ & $2{+}3{+}4$ & $3{+}5$ & $1{+}4{+}5$ \\
\end{tabular}
\end{center}

% ─── Rank-6 ──────────────────────────────────────────────────────────────────

\subsection*{Rank-6 Algebras}

\subsubsection*{C.3 (\href{https://anyonwiki.org/rings/6_1_0_16}{$\text{FR}^{6,1,0}_{16}$}) --- Categorifiable (PSU(2)$_{10}$ WZW), $D_{\text{FP}}^2 =24+12\sqrt{3}$}

\begin{center}
\small
\begin{tabular}{c|cccccc}
$\times$ & $1$ & $2$ & $3$ & $4$ & $5$ & $6$ \\
\hline
$1$ & $1$ & $2$ & $3$ & $4$ & $5$ & $6$ \\
$2$ & $2$ & $1$ & $6$ & $5$ & $4$ & $3$ \\
$3$ & $3$ & $6$ & $1{+}5{+}6$ & $4{+}5{+}6$ & $3{+}4{+}5$ & $2{+}3{+}4$ \\
$4$ & $4$ & $5$ & $4{+}5{+}6$ & $1{+}3{+}4{+}5{+}6$ & $2{+}3{+}4{+}5{+}6$ & $3{+}4{+}5$ \\
$5$ & $5$ & $4$ & $3{+}4{+}5$ & $2{+}3{+}4{+}5{+}6$ & $1{+}3{+}4{+}5{+}6$ & $4{+}5{+}6$ \\
$6$ & $6$ & $3$ & $2{+}3{+}4$ & $3{+}4{+}5$ & $4{+}5{+}6$ & $1{+}5{+}6$ \\
\end{tabular}
\end{center}

\subsubsection*{C.5 (\href{https://anyonwiki.org/rings/6_1_4_6}{$\text{FR}^{6,1,4}_{6}$}) --- Non-categorifiable, $D_{\text{FP}}^2 =12+4\sqrt{3}$}

\begin{center}
\small
\begin{tabular}{c|cccccc}
$\times$ & $1$ & $2$ & $3$ & $4$ & $5$ & $6$ \\
\hline
$1$ & $1$ & $2$ & $3$ & $4$ & $5$ & $6$ \\
$2$ & $2$ & $6$ & $1$ & $5$ & $4$ & $3$ \\
$3$ & $3$ & $1$ & $6$ & $5$ & $4$ & $2$ \\
$4$ & $4$ & $5$ & $5$ & $2{+}3{+}4{+}5$ & $1{+}4{+}5{+}6$ & $4$ \\
$5$ & $5$ & $4$ & $4$ & $1{+}4{+}5{+}6$ & $2{+}3{+}4{+}5$ & $5$ \\
$6$ & $6$ & $3$ & $2$ & $4$ & $5$ & $1$ \\
\end{tabular}
\end{center}

\subsubsection*{C.6 (\href{https://anyonwiki.org/rings/6_1_2_5}{$\text{FR}^{6,1,2}_{5}$})--- Non-categorifiable,  $D_{\text{FP}}^2 =12+4\sqrt{3}$}

\begin{center}
\small
\begin{tabular}{c|cccccc}
$\times$ & $1$ & $2$ & $3$ & $4$ & $5$ & $6$ \\
\hline
$1$ & $1$ & $2$ & $3$ & $4$ & $5$ & $6$ \\
$2$ & $2$ & $6$ & $1$ & $5$ & $4$ & $3$ \\
$3$ & $3$ & $1$ & $6$ & $5$ & $4$ & $2$ \\
$4$ & $4$ & $5$ & $5$ & $1{+}4{+}5{+}6$ & $2{+}3{+}4{+}5$ & $4$ \\
$5$ & $5$ & $4$ & $4$ & $2{+}3{+}4{+}5$ & $1{+}4{+}5{+}6$ & $5$ \\
$6$ & $6$ & $3$ & $2$ & $4$ & $5$ & $1$ \\
\end{tabular}
\end{center}

\subsubsection*{C.7 (\href{https://anyonwiki.org/rings/6_1_2_6}{$\text{FR}^{6,1,2}_{6}$})--- Non-categorifiable,  $D_{\text{FP}}^2 =12+4\sqrt{3}$}

\begin{center}
\small
\begin{tabular}{c|cccccc}
$\times$ & $1$ & $2$ & $3$ & $4$ & $5$ & $6$ \\
\hline
$1$ & $1$ & $2$ & $3$ & $4$ & $5$ & $6$ \\
$2$ & $2$ & $2{+}3{+}4{+}5$ & $1{+}2{+}3{+}6$ & $3$ & $3$ & $2$ \\
$3$ & $3$ & $1{+}2{+}3{+}6$ & $2{+}3{+}4{+}5$ & $2$ & $2$ & $3$ \\
$4$ & $4$ & $3$ & $2$ & $1$ & $6$ & $5$ \\
$5$ & $5$ & $3$ & $2$ & $6$ & $1$ & $4$ \\
$6$ & $6$ & $2$ & $3$ & $5$ & $4$ & $1$ \\
\end{tabular}
\end{center}

\subsubsection*{C.8 (\href{https://anyonwiki.org/rings/6_1_0_3}{$\text{FR}^{6,1,0}_{3}$}) --- Non-categorifiable,  $D_{\text{FP}}^2 =12+4\sqrt{3}$}

\begin{center}
\small
\begin{tabular}{c|cccccc}
$\times$ & $1$ & $2$ & $3$ & $4$ & $5$ & $6$ \\
\hline
$1$ & $1$ & $2$ & $3$ & $4$ & $5$ & $6$ \\
$2$ & $2$ & $1$ & $6$ & $5$ & $4$ & $3$ \\
$3$ & $3$ & $6$ & $1$ & $5$ & $4$ & $2$ \\
$4$ & $4$ & $5$ & $5$ & $1{+}4{+}5{+}6$ & $2{+}3{+}4{+}5$ & $4$ \\
$5$ & $5$ & $4$ & $4$ & $2{+}3{+}4{+}5$ & $1{+}4{+}5{+}6$ & $5$ \\
$6$ & $6$ & $3$ & $2$ & $4$ & $5$ & $1$ \\
\end{tabular}
\end{center}

\subsubsection*{C.9 (\href{https://anyonwiki.org/rings/6_1_4_7}{$\text{FR}^{6,1,4}_{7}$}) --- Non-categorifiable, $D_{\text{FP}}^2 =12+6\sqrt{2}$}

\begin{center}
\small
\begin{tabular}{c|cccccc}
$\times$ & $1$ & $2$ & $3$ & $4$ & $5$ & $6$ \\
\hline
$1$ & $1$ & $2$ & $3$ & $4$ & $5$ & $6$ \\
$2$ & $2$ & $3$ & $1$ & $5$ & $6$ & $4$ \\
$3$ & $3$ & $1$ & $2$ & $6$ & $4$ & $5$ \\
$4$ & $4$ & $5$ & $6$ & $3{+}4{+}6$ & $1{+}4{+}5$ & $2{+}5{+}6$ \\
$5$ & $5$ & $6$ & $4$ & $1{+}4{+}5$ & $2{+}5{+}6$ & $3{+}4{+}6$ \\
$6$ & $6$ & $4$ & $5$ & $2{+}5{+}6$ & $3{+}4{+}6$ & $1{+}4{+}5$ \\
\end{tabular}
\end{center}

\subsubsection*{C.10 (\href{https://anyonwiki.org/rings/6_1_0_10}{$\text{FR}^{6,1,0}_{10}$}) --- Non-categorifiable, $D_{\text{FP}}^2 =12+6\sqrt{2}$}

\begin{center}
\small
\begin{tabular}{c|cccccc}
$\times$ & $1$ & $2$ & $3$ & $4$ & $5$ & $6$ \\
\hline
$1$ & $1$ & $2$ & $3$ & $4$ & $5$ & $6$ \\
$2$ & $2$ & $1$ & $6$ & $4$ & $5$ & $3$ \\
$3$ & $3$ & $6$ & $1{+}4{+}5{+}6$ & $3{+}6$ & $3{+}6$ & $2{+}3{+}4{+}5$ \\
$4$ & $4$ & $4$ & $3{+}6$ & $1{+}2{+}5$ & $4{+}5$ & $3{+}6$ \\
$5$ & $5$ & $5$ & $3{+}6$ & $4{+}5$ & $1{+}2{+}4$ & $3{+}6$ \\
$6$ & $6$ & $3$ & $2{+}3{+}4{+}5$ & $3{+}6$ & $3{+}6$ & $1{+}4{+}5{+}6$ \\
\end{tabular}
\end{center}

\subsubsection*{C.11 (\href{https://anyonwiki.org/rings/6_1_4_8}{$\text{FR}^{6,1,4}_{8}$}) --- Non-categorifiable, $D_{\text{FP}}^2 =\tfrac{39}{2}+\tfrac{9}{2}\sqrt{13}$}

\begin{center}
\small
\begin{tabular}{c|cccccc}
$\times$ & $1$ & $2$ & $3$ & $4$ & $5$ & $6$ \\
\hline
$1$ & $1$ & $2$ & $3$ & $4$ & $5$ & $6$ \\
$2$ & $2$ & $3$ & $1$ & $5$ & $6$ & $4$ \\
$3$ & $3$ & $1$ & $2$ & $6$ & $4$ & $5$ \\
$4$ & $4$ & $5$ & $6$ & $3{+}4{+}5{+}6$ & $1{+}4{+}5{+}6$ & $2{+}4{+}5{+}6$ \\
$5$ & $5$ & $6$ & $4$ & $1{+}4{+}5{+}6$ & $2{+}4{+}5{+}6$ & $3{+}4{+}5{+}6$ \\
$6$ & $6$ & $4$ & $5$ & $2{+}4{+}5{+}6$ & $3{+}4{+}5{+}6$ & $1{+}4{+}5{+}6$ \\
\end{tabular}
\end{center}

\subsubsection*{C.12 (\href{https://anyonwiki.org/rings/6_1_2_10}{$\text{FR}^{6,1,2}_{10}$})--- Non-categorifiable, $D^2 \approx 36.78$}

\begin{center}
\small
\begin{tabular}{c|cccccc}
$\times$ & $1$ & $2$ & $3$ & $4$ & $5$ & $6$ \\
\hline
$1$ & $1$ & $2$ & $3$ & $4$ & $5$ & $6$ \\
$2$ & $2$ & $2{+}3{+}4{+}5$ & $1{+}2{+}3{+}6$ & $3$ & $3{+}5{+}6$ & $2{+}5{+}6$ \\
$3$ & $3$ & $1{+}2{+}3{+}6$ & $2{+}3{+}4{+}5$ & $2$ & $2{+}5{+}6$ & $3{+}5{+}6$ \\
$4$ & $4$ & $3$ & $2$ & $1$ & $6$ & $5$ \\
$5$ & $5$ & $3{+}5{+}6$ & $2{+}5{+}6$ & $6$ & $1{+}2{+}3$ & $2{+}3{+}4$ \\
$6$ & $6$ & $2{+}5{+}6$ & $3{+}5{+}6$ & $5$ & $2{+}3{+}4$ & $1{+}2{+}3$ \\
\end{tabular}
\end{center}

\subsubsection*{C.13 (\href{https://anyonwiki.org/rings/6_1_2_15}{$\text{FR}^{6,1,2}_{15}$})--- Non-categorifiable, $D^2 \approx 36.78$}

\begin{center}
\small
\begin{tabular}{c|cccccc}
$\times$ & $1$ & $2$ & $3$ & $4$ & $5$ & $6$ \\
\hline
$1$ & $1$ & $2$ & $3$ & $4$ & $5$ & $6$ \\
$2$ & $2$ & $1$ & $6$ & $5$ & $4$ & $3$ \\
$3$ & $3$ & $6$ & $1{+}4{+}5$ & $3{+}5{+}6$ & $3{+}4{+}6$ & $2{+}4{+}5$ \\
$4$ & $4$ & $5$ & $3{+}5{+}6$ & $1{+}4{+}5{+}6$ & $2{+}3{+}4{+}5$ & $3{+}4{+}6$ \\
$5$ & $5$ & $4$ & $3{+}4{+}6$ & $2{+}3{+}4{+}5$ & $1{+}4{+}5{+}6$ & $3{+}5{+}6$ \\
$6$ & $6$ & $3$ & $2{+}4{+}5$ & $3{+}4{+}6$ & $3{+}5{+}6$ & $1{+}4{+}5$ \\
\end{tabular}
\end{center}

% ─── Rank-7 ──────────────────────────────────────────────────────────────────

\subsection*{Rank-7 Algebras}

\subsubsection*{C.14 (\href{https://anyonwiki.org/rings/7_1_2_5}{$\text{FR}^{7,1,2}_{5}$}) --- Non-categorifiable, $D_{\text{FP}}^2 =17+\sqrt{17}$}

\begin{center}
\scriptsize
\renewcommand{\arraystretch}{1.15}
\begin{tabular}{c|ccccccc}
$\times$ & $1$ & $2$ & $3$ & $4$ & $5$ & $6$ & $7$ \\
\hline
$1$ & $1$ & $2$ & $3$ & $4$ & $5$ & $6$ & $7$ \\
$2$ & $2$ & $7$ & $1$ & $6$ & $5$ & $4$ & $3$ \\
$3$ & $3$ & $1$ & $7$ & $6$ & $5$ & $4$ & $2$ \\
$4$ & $4$ & $6$ & $6$ & $1{+}5{+}6{+}7$ & $4{+}6$ & $2{+}3{+}4{+}5$ & $4$ \\
$5$ & $5$ & $5$ & $5$ & $4{+}6$ & $1{+}2{+}3{+}7$ & $4{+}6$ & $5$ \\
$6$ & $6$ & $4$ & $4$ & $2{+}3{+}4{+}5$ & $4{+}6$ & $1{+}5{+}6{+}7$ & $6$ \\
$7$ & $7$ & $3$ & $2$ & $4$ & $5$ & $6$ & $1$ \\
\end{tabular}
\end{center}

\subsubsection*{C.15 (\href{https://anyonwiki.org/rings/7_1_0_2}{$\text{FR}^{7,1,0}_{2}$}) --- Non-categorifiable, $D_{\text{FP}}^2 =17+\sqrt{17}$}

\begin{center}
\scriptsize
\renewcommand{\arraystretch}{1.15}
\begin{tabular}{c|ccccccc}
$\times$ & $1$ & $2$ & $3$ & $4$ & $5$ & $6$ & $7$ \\
\hline
$1$ & $1$ & $2$ & $3$ & $4$ & $5$ & $6$ & $7$ \\
$2$ & $2$ & $1$ & $7$ & $6$ & $5$ & $4$ & $3$ \\
$3$ & $3$ & $7$ & $1$ & $6$ & $5$ & $4$ & $2$ \\
$4$ & $4$ & $6$ & $6$ & $1{+}5{+}6{+}7$ & $4{+}6$ & $2{+}3{+}4{+}5$ & $4$ \\
$5$ & $5$ & $5$ & $5$ & $4{+}6$ & $1{+}2{+}3{+}7$ & $4{+}6$ & $5$ \\
$6$ & $6$ & $4$ & $4$ & $2{+}3{+}4{+}5$ & $4{+}6$ & $1{+}5{+}6{+}7$ & $6$ \\
$7$ & $7$ & $3$ & $2$ & $4$ & $5$ & $6$ & $1$ \\
\end{tabular}
\end{center}

\subsubsection*{C.16 (\href{https://anyonwiki.org/rings/7_1_0_11}{$\text{FR}^{7,1,0}_{11}$}) --- Non-categorifiable, $D_{\text{FP}}^2 =20+12\sqrt{2}$}

\begin{center}
\scriptsize
\renewcommand{\arraystretch}{1.15}
\begin{tabular}{c|ccccccc}
$\times$ & $1$ & $2$ & $3$ & $4$ & $5$ & $6$ & $7$ \\
\hline
$1$ & $1$ & $2$ & $3$ & $4$ & $5$ & $6$ & $7$ \\
$2$ & $2$ & $1$ & $7$ & $6$ & $5$ & $4$ & $3$ \\
$3$ & $3$ & $7$ & $1{+}6{+}7$ & $5{+}7$ & $4{+}5{+}6$ & $3{+}5$ & $2{+}3{+}4$ \\
$4$ & $4$ & $6$ & $5{+}7$ & $1{+}4{+}6$ & $3{+}5{+}7$ & $2{+}4{+}6$ & $3{+}5$ \\
$5$ & $5$ & $5$ & $4{+}5{+}6$ & $3{+}5{+}7$ & $1{+}2{+}3{+}4{+}6{+}7$ & $3{+}5{+}7$ & $4{+}5{+}6$ \\
$6$ & $6$ & $4$ & $3{+}5$ & $2{+}4{+}6$ & $3{+}5{+}7$ & $1{+}4{+}6$ & $5{+}7$ \\
$7$ & $7$ & $3$ & $2{+}3{+}4$ & $3{+}5$ & $4{+}5{+}6$ & $5{+}7$ & $1{+}6{+}7$ \\
\end{tabular}
\end{center}

\subsubsection*{C.17 (\href{https://anyonwiki.org/rings/7_1_0_12}{$\text{FR}^{7,1,0}_{12}$}) --- Non-categorifiable, $D_{\text{FP}}^2= 42$}

\begin{center}
\scriptsize
\renewcommand{\arraystretch}{1.15}
\begin{tabular}{c|ccccccc}
$\times$ & $1$ & $2$ & $3$ & $4$ & $5$ & $6$ & $7$ \\
\hline
$1$ & $1$ & $2$ & $3$ & $4$ & $5$ & $6$ & $7$ \\
$2$ & $2$ & $1$ & $7$ & $6$ & $5$ & $4$ & $3$ \\
$3$ & $3$ & $7$ & $1{+}5{+}6{+}7$ & $4{+}6{+}7$ & $3{+}7$ & $3{+}4{+}6$ & $2{+}3{+}4{+}5$ \\
$4$ & $4$ & $6$ & $4{+}6{+}7$ & $1{+}3{+}5{+}7$ & $4{+}6$ & $2{+}3{+}5{+}7$ & $3{+}4{+}6$ \\
$5$ & $5$ & $5$ & $3{+}7$ & $4{+}6$ & $1{+}2{+}5$ & $4{+}6$ & $3{+}7$ \\
$6$ & $6$ & $4$ & $3{+}4{+}6$ & $2{+}3{+}5{+}7$ & $4{+}6$ & $1{+}3{+}5{+}7$ & $4{+}6{+}7$ \\
$7$ & $7$ & $3$ & $2{+}3{+}4{+}5$ & $3{+}4{+}6$ & $3{+}7$ & $4{+}6{+}7$ & $1{+}5{+}6{+}7$ \\
\end{tabular}
\end{center}

\subsubsection*{C.18  (\href{https://anyonwiki.org/rings/7_1_0_9}{$\text{FR}^{7,1,0}_{9}$}) --- Non-categorifiable, $D_{\text{FP}}^2= 29+\sqrt{29}$}

\begin{center}
\scriptsize
\renewcommand{\arraystretch}{1.15}
\begin{tabular}{c|ccccccc}
$\times$ & $1$ & $2$ & $3$ & $4$ & $5$ & $6$ & $7$ \\
\hline
$1$ & $1$ & $2$ & $3$ & $4$ & $5$ & $6$ & $7$ \\
$2$ & $2$ & $1$ & $7$ & $4$ & $5$ & $6$ & $3$ \\
$3$ & $3$ & $7$ & $1{+}4{+}5{+}6{+}7$ & $3{+}7$ & $3{+}7$ & $3{+}7$ & $2{+}3{+}4{+}5{+}6$ \\
$4$ & $4$ & $4$ & $3{+}7$ & $1{+}2{+}6$ & $5{+}6$ & $4{+}5$ & $3{+}7$ \\
$5$ & $5$ & $5$ & $3{+}7$ & $5{+}6$ & $1{+}2{+}4$ & $4{+}6$ & $3{+}7$ \\
$6$ & $6$ & $6$ & $3{+}7$ & $4{+}5$ & $4{+}6$ & $1{+}2{+}5$ & $3{+}7$ \\
$7$ & $7$ & $3$ & $2{+}3{+}4{+}5{+}6$ & $3{+}7$ & $3{+}7$ & $3{+}7$ & $1{+}4{+}5{+}6{+}7$ \\
\end{tabular}
\end{center}

\subsection*{Rank-8 Algebra}

\subsubsection*{\href{https://anyonwiki.org/rings/8_1_6_8}{$\mathrm{FR}^{8,6}_{8}$} --- Non-categorifiable, $D_{\text{FP}}^2= 16+8\sqrt{2}$,  $\rm{cl}(\A)=3$}

\begin{center}
\scriptsize
\renewcommand{\arraystretch}{1.15}
\begin{tabular}{c|cccccccc}
$\times$ & $1$ & $2$ & $3$ & $4$ & $5$ & $6$ & $7$ & $8$ \\
\hline
$1$ & $1$ & $2$ & $3$ & $4$ & $5$ & $6$ & $7$ & $8$ \\
$2$ & $2$ & $1$ & $4$ & $3$ & $8$ & $7$ & $6$ & $5$ \\
$3$ & $3$ & $4$ & $2$ & $1$ & $6$ & $8$ & $5$ & $7$ \\
$4$ & $4$ & $3$ & $1$ & $2$ & $7$ & $5$ & $8$ & $6$ \\
$5$ & $5$ & $8$ & $6$ & $7$ & $4{+}6{+}8$ & $1{+}7{+}8$ & $2{+}5{+}6$ & $3{+}5{+}7$ \\
$6$ & $6$ & $7$ & $8$ & $5$ & $1{+}7{+}8$ & $3{+}5{+}7$ & $4{+}6{+}8$ & $2{+}5{+}6$ \\
$7$ & $7$ & $6$ & $5$ & $8$ & $2{+}5{+}6$ & $4{+}6{+}8$ & $3{+}5{+}7$ & $1{+}7{+}8$ \\
$8$ & $8$ & $5$ & $7$ & $6$ & $3{+}5{+}7$ & $2{+}5{+}6$ & $1{+}7{+}8$ & $4{+}6{+}8$
\end{tabular}
\end{center}

% ============================================================
\section{The $\rm{cl}(\A)=2$ Analysis in Detail}
\label{app:cl2}
% ============================================================

This appendix collects the full analysis of the multiplicity-free fusion
algebras of rank $\leq7$ which have $\mathrm{cl}(\A)=2$.  The results
are summarised in Sec.~\ref{sec:cl2}; the fusion tables are in
Appendix~\ref{app:classification}.  Throughout, scores are computed with the
metrics of Sec.~\ref{sec:metrics} and assignments are ranked by ensemble medians. The loop-order dictionary is that of
Sec.~\ref{sec:yukawa-setup} truncated at $\mathrm{cl}(\A)=2$. 

Throughout, the assignments are ranked by the median of the relevant score over the ensemble of Eq.~\eqref{eq:ensemble}.

\subsection{Charged leptons: scan over all algebras}
\label{app:cl2-lepton-scan}

In order to find the best possible hierarchy in each algebra listed in
Appendix~\ref{app:classification}, we assign the basis elements to the fields
in all possible ways, compute the resulting mass ratios, and select the
assignment minimising the metric $\rho$ of Eq.~\eqref{eq:rho}. 
\begin{center}\small
\renewcommand{\arraystretch}{1.15}
\begin{tabular}{|c|c|c|c|}
\hline
Algebra & $1:\tfrac{m_\mu}{m_e}:\tfrac{m_\tau}{m_e}$ (MC median)
        & $\rho_\ell$ median & $68\%$ band \\
\hline
C.1a, C.1b, C.2, C.3 & $1:1394:2641$ & $1.122$ & $[0.842,1.820]$ \\
C.5, C.6, C.7, C.8 & $1:101.2:3851$ & $0.723$ & $[0.392,1.228]$ \\
\textbf{C.9} & $1:101.2:3847$ & $\mathbf{0.722}$ & $[0.393,1.229]$ \\
C.10 & $1:1394:2641$ & $1.122$ & $[0.842,1.820]$ \\
C.11 & $1:68.8:2805$ & $0.880$ & $[0.468,1.649]$ \\
C.12, C.13 & $1:79.4:3191$ & $0.812$ & $[0.438,1.491]$ \\
C.14, C.15 & $1:105.5:6218$ & $0.734$ & $[0.466,1.162]$ \\
C.16 & $1:78.3:4958$ & $0.824$ & $[0.507,1.420]$ \\
C.17 & $1:934.7:3250$ & $0.930$ & $[0.598,1.822]$ \\
C.18 & $1:1394:2641$ & $1.122$ & $[0.842,1.820]$ \\
\hline
Observed & $1:210.7:3578$ & & \\
\hline
\end{tabular}
\end{center}

We see that there are multiple algebras that give a pretty good fit to the observed mass ratios. However, when we also consider the quark sector and the neutrino sector, we will see that the available algebras get limited.

\subsection{Neutrinos}
\label{app:cl2-neutrino}

\subsubsection*{The C.9 neutrino texture}
 With the specific example we considered in the previous section, we have $H^*[1]$ and $l_1[4],l_2[1],l_3[3]$. This gives the following mass matrix for the neutrinos
 \begin{equation}
    M_\nu \sim \frac{v^2}{\Lambda}
    \begin{pmatrix}
      \eps   & \eps   & \eps^2\\
      \eps   & 1      & \eps^2\\
      \eps^2 & \eps^2 & \eps^2
    \end{pmatrix}
    \label{eq:nu-texture}
\end{equation}
The singular values of this matrix are in the ratio
\begin{equation}    
    m_{\nu_1}:m_{\nu_2}:m_{\nu_3} \approx \eps^2:\eps:1
\end{equation}
which gives a hierarchy of neutrino masses.

In order to check with observations, we need to fix the scale $\Lambda$. We do this by fixing the largest neutrino mass to be $m_{\nu_3} = 0.05$ eV (the atmospheric neutrino mass scale). Given this, we find that the scale $\Lambda$ is given by typical GUT scale, $\Lambda \approx 6.17\times10^{14}$ GeV. In the case of the example above, we find that 
\begin{equation}
    \Delta m^2_{31} \approx m_3^2 \approx 2.5\times10^{-3} \text{ eV}^2
\end{equation}
which is consistent with the observed value. But, note that this will be the case by construction as long as there is a hierarchy in the neutrino masses in our setup since we are fixing the largest neutrino mass to be $0.05$ eV. 

\subsubsection*{Scan over all algebras}
As before we scan over all the algebras, in each case evaluating the
Weinberg texture at that algebra's best charged-lepton assignment and quoting
the median of the ensemble of Eq.~\eqref{eq:ensemble} with symmetric
coefficients.  The charged-lepton score is repeated in the second column.

\begin{table}[H]
\centering\small
\renewcommand{\arraystretch}{1.2}
\begin{tabular}{|c|c|c|c|c|}
\hline
Algebra & $\rho_\ell$ & $m_{\nu_3}:m_{\nu_2}:m_{\nu_1}$
        & $\tfrac{\Delta m^2_{21}}{\Delta m^2_{31}}$ & $68\%$ band \\
\hline
C.1a & $1.122$ & $1:0.946:0.57$ & $0.48$ & $[0.012,0.97]$ \\
C.1b, C.2, C.3 & $1.122$ & $1:0.686:0.40$ & $0.29$ & $[0.067,0.70]$ \\
 C.11 & $0.880$ & $1:0.0301:0.010$ & $7.0\times10^{-4}$ & $[1.8\times10^{-4},2.8\times10^{-3}]$ \\
C.5, C.7 & $0.723$ & $1:0.585:0.012$ & $0.34$ & $[0.11,0.73]$ \\
C.6, C.8 & $0.723$ & $1:0.686:0.40$ & $0.29$ & $[0.066,0.70]$ \\
\textbf{C.9} & $\mathbf{0.722}$ & $1:0.0170:2.9\times10^{-4}$ & $\mathbf{2.9\times10^{-4}}$ & $[5.8\times10^{-5},1.4\times10^{-3}]$ \\
C.10, C.18 & $1.122$ & $1:0.686:0.40$ & $0.29$ & $[0.067,0.70]$ \\
C.12 & $0.812$ & $1:0.585:0.012$ & $0.34$ & $[0.11,0.73]$ \\
C.13 & $0.812$ & $1:0.687:0.40$ & $0.29$ & $[0.067,0.70]$ \\
C.14, C.15 & $0.734$ & $1:0.632:0.28$ & $0.28$ & $[0.060,0.70]$ \\
C.16 & $0.824$ & $1:0.665:0.22$ & $0.37$ & $[0.097,0.74]$ \\
C.17 & $0.930$ & $1:0.633:0.23$ & $0.34$ & $[0.10,0.73]$ \\
\hline
Observed & & --- & $3.0\times10^{-2}$ & \\
\hline
\end{tabular}
\caption{Neutrino spectra for the algebras that survive the charged-lepton
fit, each evaluated at its best-fit left-handed lepton assignment.  For every algebra the scale is fixed by hand
through $m_{\nu_3}=0.05$~eV, which sets $\Lambda\approx6\times10^{14}$~GeV.
}
\label{tab:nu_scan}
\end{table}

\paragraph{A comment on the PMNS matrix:}
Unlike the CKM matrix, the observed PMNS matrix has no hierarchial structure, so any assignment that predicts a hierarchial PMNS matrix is disfavored. Hierarchial textures for $Y^e$ and $M_\nu$ generally produce such a hierarchy. In particular we note here that the PMNS matrix for the C.9 assignment does not match the observed PMNS matrix. If this was the algebra that was realized in nature, then there must be additional contributions to the PMNS matrix from the neutrino sector.

\subsection{Quarks}
\label{sec:quark}

\subsubsection*{Down-type Yukawa texture}
Going forward with the algebra C.9, we assign for the down-type quarks $d_j^c$, a similar assignment as the leptons $e_j^c$ and $q_i$, a similar assignment as $l_i$. That is, 
\begin{equation}
    H^*[1],\quad q_1[4],q_2[1],q_3[3],\quad d^c_1[6],d^c_2[2],d^c_3[3]
\end{equation}
Therefore, the resulting down-type Yukawa matrix will have the same texture as the lepton Yukawa matrix. 
\begin{equation}
    Y^d \sim \begin{pmatrix}
      \eps   & \eps   & \eps^2\\
      \eps^2 & \eps^2 & \eps^2\\
      \eps   & 1      & \eps^2
    \end{pmatrix},
    \qquad
    m_d:m_s:m_b \approx 1:58.8:3522
    \quad \nw{\left( \sim 1:\eps^{-1}:\eps^{-2} \right)},
\end{equation}
against the observed $1:19.9:1058$. This is a pretty good fit to the observed mass ratios, upto $\mathcal{O}(1)$ Wilson coefficient corrections.

\subsubsection*{Up-type Yukawa texture}
The up-type quarks have the largest hierarchy in the Standard Model. Therefore, we need to find an assignment that gives a large hierarchy. For the specific algebra C.9, with the choice of assignments for $q_i$ and $H$ as before, we find that the following assignment for the right-handed up-type quarks gives a good fit to the observed mass ratios:
\begin{equation}
u^c_1[4],u^c_2[3],u^c_3[2]
\end{equation}
giving the up-type Yukawa matrix the texture
\begin{equation}
    Y^u \sim \begin{pmatrix}
      \eps   & \eps^2 & \eps\\
      \eps   & \eps^2 & \eps^2\\
      \eps^2 & \eps^2 & 1
    \end{pmatrix}
\end{equation}
giving the singular values in the ratio
\begin{equation}
    m_u:m_c:m_t \approx 1:7162:297956
    \quad \left( \sim 1: \eps^{-2}: \eps^{-3} \right),
\end{equation}
against the observed $1:506:137400$. This assignment is the only one in our scan that reaches the
third power $\eps^{-3}$, and that extra power is what makes the very
large top-quark hierarchy accessible at all. But, it does so only at
$c_{ij}=1$.  Under the ensemble of Eq.~\eqref{eq:ensemble} the median
ratios collapse to $1:100.7:3854$, for the reason set out in
Sec.~\ref{sec:cl2-failure}. The leading term of $\det Y^u$ vanishes identically
at unit coefficients and is restored by any generic choice.

\subsubsection*{Scan over all algebras}
Repeating the same procedure as in the lepton sector, we scan over all the algebras and find the best fit for the up-type and down-type quark mass ratios. We find following best fits for both the up-type and down-type quarks
\begin{table}[H]
\centering\small
\renewcommand{\arraystretch}{1.15}
\begin{tabular}{|l|c|c|c|}
\hline
Algebra & $1:\tfrac{m_c}{m_u}:\tfrac{m_t}{m_u}$ (MC median)
        & $\rho_u$ median & $68\%$ band \\
\hline
C.1a, C.1b, C.2, C.3, C.10, C.18
    & $1:1394:2641$ & $2.247$ & $[2.004,2.408]$ \\
 C.11
    & $1:1849:3521$ & $2.225$ & $[1.981,2.389]$ \\
C.5, C.6, C.7, C.8
    & $1:2541:4729$ & $2.200$ & $[1.956,2.367]$ \\
\textbf{C.9}, C.12, C.13, C.17
    & $1:934.7:3250$ & $2.046$ & $[1.784,2.443]$ \\
C.14, C.15, C.16
    & $1:1954:5735$ & $\mathbf{2.004}$ & $[1.756,2.219]$ \\
\hline
Observed & $1:506:137400$ & --- & \\
\hline
\end{tabular}
\caption{Best up-type fit per algebra with the corresponding score.  }
\label{tab:up_scan}
\end{table}

\begin{table}[H]
\centering\small
\renewcommand{\arraystretch}{1.15}
\begin{tabular}{|l|c|c|c|}
\hline
Algebra & $1:\tfrac{m_s}{m_d}:\tfrac{m_b}{m_d}$ (MC median)
        & $\rho_d$ median & $68\%$ band \\
\hline
\textbf{C.9}
    & $1:42.3:2740$ & $\mathbf{0.803}$ & $[0.362,1.853]$ \\
C.12, C.13
    & $1:42.2:2736$ & $0.804$ & $[0.363,1.847]$ \\
C.5--C.8, C.11, C.14, C.15, C.16
    & $1:68.8:2805$ & $0.971$ & $[0.443,2.034]$ \\
C.2, C.10, C.17, C.18
    & $1:65.0:208.2$ & $1.268$ & $[1.021,1.476]$ \\
C.1a, C.1b, C.3
    & $1:31.1:113.7$ & $1.289$ & $[1.038,1.513]$ \\
\hline
Observed & $1:19.9:1058$ & --- & \\
\hline
\end{tabular}
\caption{Best down-type fit per algebra with the corresponding score. }
\label{tab:down_scan}
\end{table}

\subsection{The CKM matrix}
\label{sec:CKM}

The CKM matrix is constructed from $Y^u$ and $Y^d$ as described in
Sec.~\ref{sec:rank8-ckm}.

\subsubsection*{The C.9 example}
Let us consider the algebra C.9 with the assignments we have chosen for the quarks, namely $H^*[1]$, $q_1[4],q_2[1],q_3[3]$, $u^c_1[4],u^c_2[3],u^c_3[2]$ and $d^c_1[6],d^c_2[2],d^c_3[3]$. Recall that the resulting Yukawa matrices are
\begin{equation}
    Y^u \sim \begin{pmatrix}
      \eps   & \eps^2 & \eps\\
      \eps   & \eps^2 & \eps^2\\
      \eps^2 & \eps^2 & 1
    \end{pmatrix},
    \qquad
    Y^d \sim \begin{pmatrix}
      \eps   & \eps   & \eps^2\\
      \eps^2 & \eps^2 & \eps^2\\
      \eps   & 1      & \eps^2
    \end{pmatrix}.
\end{equation}
Diagonalizing these matrices, and using the best fit value of $\eps=0.017$, we find that the CKM matrix is given by
\begin{equation}
 |V_{\rm CKM}^{\text{C.9}}| \approx
 \begin{pmatrix}
 0.719 & 0.695 & 0.0002\\
 0.695 & 0.719 & 0.0002\\
 0.00001 & 0.0003 & 1.000
 \end{pmatrix}.
 \label{eq:ckm-c9}
\end{equation}
\nw{Expanding the same diagonalization to leading order in $\eps$ instead, the
texture is}
\begin{equation}
 |V_{\rm CKM}^{\text{C.9}}| \sim
 \begin{pmatrix}
 \mathcal{O}(1) & \mathcal{O}(1) & \mathcal{O}(\eps^2)\\
 \mathcal{O}(1) & \mathcal{O}(1) & \mathcal{O}(\eps^2)\\
 \mathcal{O}(\eps^3) & \mathcal{O}(\eps^2) & 1
 \end{pmatrix},
 \label{eq:ckm-c9-eps}
\end{equation}
which carries information the decimals hide.  The upper-left $\mathcal{O}(1)$
block is not accidentally large: it is a near-maximal rotation, $|V_{us}|\sim\sin\theta_{12}$
with $\theta_{12}\approx43^\circ$, and no choice of $\mathcal{O}(1)$ Wilson
coefficients of natural size will bring it down to the observed Cabibbo angle
without a tuning.  By contrast the smallness of the third-generation mixings is
parametric, and the numerically tiny $|V_{td}|\approx10^{-5}$ is
$\mathcal{O}(\eps^3)$ .

To compare with, recall the observed CKM matrix of Eq.~\eqref{eq:ckm-obs}.
We see that some of the qualitative features of the CKM matrix are captured by the fusion algebra structure, such as the fact that the third generation is mostly decoupled from the first two generations. Furthermore, the first two generations are strongly mixed with each other. However, the mixing between the first and second generation is too large compared to the observed value. These discrepancies could be fixed by adding $\mathcal{O}(1)$ Wilson coefficient corrections to the Yukawa matrices.

\subsubsection*{Scan over algebras}

Using $\rho_{\rm CKM}$ of Eq.~\eqref{eq:rho-ckm} and the combined mass score
$\rho_{\rm mass}$ of Eq.~\eqref{eq:rho-mass}, we scan over all the algebras and
find the following.  First, the assignment that best fits the quark mass ratios
as given in Sec.~\ref{sec:quark} is not the best fit for the CKM matrix. The
mass-optimal assignment gives median scores $\rho_{\rm CKM}=0.444$ and
$\rho_{\rm mass}=1.79$.  Second, the assignment that best fits the CKM
matrix gives $\rho_{\rm CKM}=0.316$ with $\rho_{\rm mass}=2.19$.
This is achieved for the same algebra \textbf{C.9}, now with the
assignments $H^*[1]$, $q_i=(1,2,5)$, $u^c_i=(1,2,5)$, $d^c_i=(1,5,6)$. For
reference, the assignment quoted below is the one that minimised
$\rho_{\rm CKM}$ at unit Wilson coefficients,
\begin{equation}
    H[4],\quad q_i=[4,1,2],\quad u^c_i=[1,6,5],\quad d^c_i=[1,2,6],
\end{equation}
giving
\begin{equation}
 |V_{\rm CKM}| \approx
 \begin{pmatrix}
 0.800 & 0.600 & 0.0001\\
 0.600 & 0.800 & 0.0257\\
 0.016 & 0.021 & 0.9997
 \end{pmatrix},
 \qquad \rho_{\rm CKM} = 0.36 ,
\end{equation}
\nw{or, to leading order in $\eps$,}
\nw{\begin{equation}
 |V_{\rm CKM}| \sim
 \begin{pmatrix}
 \mathcal{O}(1) & \mathcal{O}(1) & \mathcal{O}(\eps^2)\\
 \mathcal{O}(1) & \mathcal{O}(1) & \mathcal{O}(\eps)\\
 \mathcal{O}(\eps) & \mathcal{O}(\eps) & 1
 \end{pmatrix}.
 \label{eq:ckm-best-eps}
\end{equation}}
\nw{Comparing with the mass-optimal texture of Eq.~\eqref{eq:ckm-c9-eps} shows
exactly what the CKM-optimal assignment is trading away.  The
third-generation mixings are enhanced by one power of $\eps$ ---
$|V_{cb}|,|V_{td}|,|V_{ts}|\sim\eps$ rather than $\eps^2$ or $\eps^3$ --- which
is what brings them close to the observed $|V_{cb}|\approx0.042$ and
$|V_{ts}|\approx0.041$, at the cost of a considerably worse mass fit
($\rho_{\rm mass}$ degrades from $1.26$ to $1.95$).}
The best fits for each algebra are summarized in appendix~\ref{sec:ckm_scan}.

\subsection{CKM Best-Fit Parameters for Each Algebra}
\label{sec:ckm_scan}

\begin{table}[H]
\centering\small
\renewcommand{\arraystretch}{1.2}
\begin{tabular}{|l|c|c|c|c|c|c|}
\hline
Algebra & $H$ & $q_i$ & $u^c_i$ & $d^c_i$ & $\rho_{\rm mass}$ & $\rho_{\rm CKM}$ \\
\hline
C.1a & $2$ & $(1,2,4)$ & $(1,2,4)$ & $(1,3,4)$ & 3.180 & 0.600 \\
C.1b & $3$ & $(1,2,3)$ & $(1,2,3)$ & $(1,2,3)$ & 3.183 & 0.598 \\
C.2  & $1$ & $(1,2,4)$ & $(1,3,5)$ & $(1,3,5)$ & 3.911 & 0.389 \\
C.3  & $1$ & $(1,3,4)$ & $(2,3,5)$ & $(2,3,5)$ & 3.909 & 0.390 \\
C.5  & $1$ & $(1,4,6)$ & $(1,2,3)$ & $(1,2,4)$ & 2.040 & 0.335 \\
C.6  & $1$ & $(1,4,6)$ & $(1,2,3)$ & $(1,2,5)$ & 2.045 & 0.336 \\
C.7  & $1$ & $(1,2,6)$ & $(1,2,4)$ & $(1,4,5)$ & 3.871 & 0.334 \\
C.8  & $1$ & $(1,4,5)$ & $(2,3,4)$ & $(2,4,6)$ & 2.214 & 0.335 \\
\textbf{C.9} & $1$ & $(1,2,5)$ & $(1,2,5)$ & $(1,5,6)$ & 2.196 & \textbf{0.316} \\
C.10 & $1$ & $(1,2,4)$ & $(1,5,6)$ & $(1,3,6)$ & 3.909 & 0.389 \\
C.11 & $1$ & $(1,4,5)$ & $(2,3,4)$ & $(2,4,6)$ & 2.214 & 0.335 \\
C.12 & $1$ & $(1,2,3)$ & $(2,4,5)$ & $(2,4,6)$ & 2.217 & 0.335 \\
C.13 & $1$ & $(1,4,6)$ & $(2,3,6)$ & $(3,5,6)$ & 2.149 & 0.334 \\
C.14 & $1$ & $(1,5,6)$ & $(1,2,3)$ & $(1,2,7)$ & 2.218 & 0.335 \\
C.15 & $1$ & $(1,5,6)$ & $(1,2,7)$ & $(1,2,3)$ & 3.837 & 0.335 \\
C.16 & $1$ & $(3,4,5)$ & $(1,3,6)$ & $(1,2,3)$ & 3.831 & 0.335 \\
C.17 & $1$ & $(1,2,3)$ & $(1,5,6)$ & $(1,5,6)$ & 3.905 & 0.390 \\
C.18 & $1$ & $(1,2,4)$ & $(1,5,7)$ & $(1,5,7)$ & 3.909 & 0.389 \\
\hline
\end{tabular}
\caption{Best CKM fit per algebra under the ensemble of
Eq.~\eqref{eq:ensemble}.  For each algebra we list the assignment
$(H,q_i,u^c_i,d^c_i)$ minimising the \emph{median} of $\rho_{\rm CKM}$,
together with the median combined mass score $\rho_{\rm mass}$ at that
assignment.  }
\label{tab:ckm_scan}
\end{table}
The spread across algebras is small --- twelve of the nineteen
lie between $0.31$ and $0.34$ --- so the CKM matrix on its own does very little
to discriminate between them. What discriminates is the mass sector, and every
CKM-optimal assignment has a poor $\rho_{\rm mass}$ with the best fit coming from the C.9 algebra .

\section{Feynman Diagrams for the C.9 lepton texture}
In this appendix we present the Feynman diagrams corresponding to the C.9 lepton texture. Note that the fusion rules can be easily interpreted in a diagrammatic way as follows.
\begin{itemize}
    \item Three lines can meet at a vertex if the corresponding fusion rule gives identity.
    \item A line in the opposite direction corresponds to the conjugate representation.\footnote{Note that the arrowheads correspond to the direction of the flow of the representation, not the flow of charge. Self conjugate fields are denoted without an arrowhead since they are insensitive to the ``direction'' in which they fuse.}
\end{itemize}
A point to note is that there are multiple ways to fuse the lines together. Here we show one possible way to do so, but there are other ways to do it as well. The assignments for which the diagrams below contribute is 
\begin{equation}
    \bar{H}=1\qquad,\qquad l_i=\{4,1,3\}\qquad,\qquad e^c_i=\{6,2,3\}
\end{equation}
In drawing the figures, we have shown one particular example where a choice on the Lorentz spin of the fields running in the loops has been made. Dashed lines represent scalars and bold lines represent fermions.

\newcommand{\CNineLineStyle}[1]{%
  \ifnum#1=6
  \else
    fermion
  \fi
}

\newcommand{\CNineLineStyleScalar}[1]{
  \ifnum #1=6
  scalar
  \else
     charged scalar
  \fi
}

% ==========================================================================
% Tree-level diagram
% ==========================================================================

\newcommand{\CNineTreeDiagram}[3]{%
\begin{tikzpicture}
  \begin{feynman}
    \vertex (v) at (0,0);

    \vertex (e1) at (90:2)  {$\bar{H}[1]$};
    \vertex (e2) at (210:2) {$#1[#2]$};
    \vertex (e3) at (330:2) {$#3$};

    \diagram* {
      (e1) -- [charged scalar] (v),
      (e2) -- [\CNineLineStyle{#2}] (v),
      (e3) -- [fermion] (v),
    };
  \end{feynman}
\end{tikzpicture}
}

% ==========================================================================
% One-loop diagram
% ==========================================================================

\newcommand{\CNineOneLoopDiagram}[4]{%
\begin{tikzpicture}
  \begin{feynman}
    \vertex (a) at (90:1.5);
    \vertex (b) at (210:1.5);
    \vertex (c) at (330:1.5);

    \vertex (e1) at (90:3)  {$\bar{H}[1]$};
    \vertex (e2) at (210:3) {$#1[#2]$};
    \vertex (e3) at (330:3) {$#3[#4]$};

    \diagram* {
      (e1) -- [charged scalar] (a),
      (e2) -- [\CNineLineStyle{#2}] (b),
      (e3) -- [\CNineLineStyle{#4}] (c),

      (a) -- [
        anti fermion,
        bend left=60,
        edge label=$4$
      ] (c),

      (c) -- [scalar,
        bend left=60,
        edge label=$6$
      ] (b),

      (b) -- [
        fermion,
        bend left=60,
        edge label=$5$
      ] (a),
    };
  \end{feynman}
\end{tikzpicture}
}

% ==========================================================================
% Two-loop diagram
% ==========================================================================

\newcommand{\CNineTwoLoopDiagram}[8]{
\begin{tikzpicture}
  \begin{feynman}
    \vertex (a) at (0,1.6);
    \vertex (b) at (-2.2,-1.0);
    \vertex (d) at (-0.8,-1.0);
    \vertex (e) at (0.8,-1.0);
    \vertex (c) at (2.2,-1.0);

    \vertex (x1) at (0,3.4)     {$\bar{H}[1]$};
    \vertex (x2) at (-3.8,-2.0) {$#1[#2]$};
    \vertex (x3) at (3.8,-2.0)  {$#3[#4]$};

    \diagram* {
      (x1) -- [charged scalar] (a),

      (x2) -- [
        \CNineLineStyle{#2}
      ] (b),

      (x3) -- [
        \CNineLineStyle{#4}
      ] (c),

      (a) -- [
        \CNineLineStyle{#5},
        edge label=$#5$
      ] (b),

      (b) -- [
        \CNineLineStyleScalar{#6},
        edge label'=$#6$
      ] (d),

      (d) -- [
        fermion,
        bend left=75,
        edge label=$4$
      ] (e),

      (d) -- [
        fermion,
        bend right=75,
        edge label'=$4$
      ] (e),

      (e) -- [
        \CNineLineStyleScalar{#7},
        edge label'=$#7$
      ] (c),

      (c) -- [
        \CNineLineStyle{#8},
        edge label=$#8$
      ] (a),
    };
  \end{feynman}
\end{tikzpicture}%
}

% ==========================================================================
\subsection{Tree-level contribution}
% ==========================================================================

The coupling $\bar{H}l_3e_2^c$ is allowed directly by the fusion rules and
therefore appears at tree level.

\begin{figure}[H]
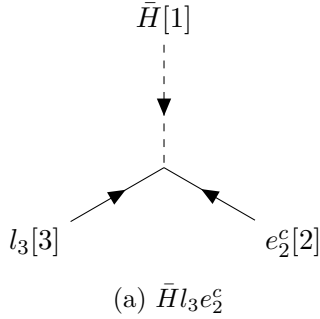

    \centering

    \begin{subfigure}[t]{0.42\textwidth}
        \centering
        \resizebox{0.75\linewidth}{!}{
            \CNineTreeDiagram{l_3}{3}{e_2^c[2]}
        }
        \caption{$\bar{H}l_3e_2^c$}
        \label{fig:C9-tree-l3e2}
    \end{subfigure}

    \caption{Tree-level realization of the C.9 charged-lepton texture.}
    \label{fig:C9-tree}
\end{figure}

% ==========================================================================
\subsection{One-loop contributions}
% ==========================================================================

\begin{figure}[H]
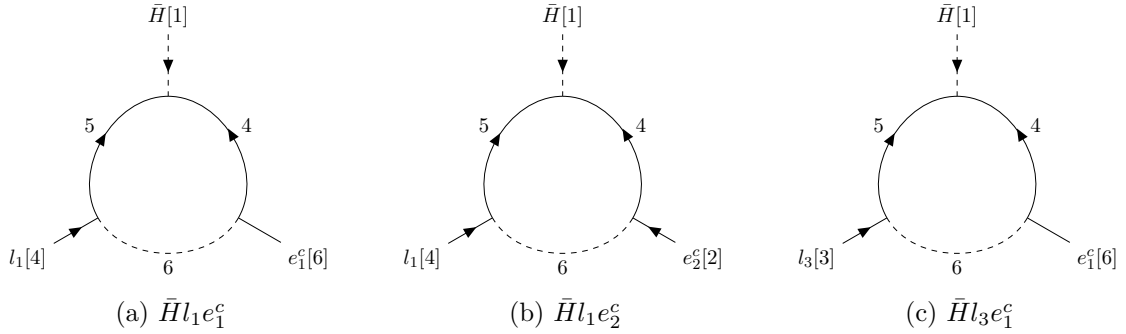

    \centering

    \begin{subfigure}[t]{0.31\textwidth}
        \centering
        \resizebox{\linewidth}{!}{
            \CNineOneLoopDiagram{l_1}{4}{e_1^c}{6}
        }
        \caption{$\bar{H}l_1e_1^c$}
        \label{fig:C9-one-loop-l1e1}
    \end{subfigure}
    \hfill
    \begin{subfigure}[t]{0.31\textwidth}
        \centering
        \resizebox{\linewidth}{!}{
            \CNineOneLoopDiagram{l_1}{4}{e_2^c}{2}
        }
        \caption{$\bar{H}l_1e_2^c$}
        \label{fig:C9-one-loop-l1e2}
    \end{subfigure}
    \hfill
    \begin{subfigure}[t]{0.31\textwidth}
        \centering
        \resizebox{\linewidth}{!}{
            \CNineOneLoopDiagram{l_3}{3}{e_1^c}{6}
        }
        \caption{$\bar{H}l_3e_1^c$}
        \label{fig:C9-one-loop-l3e1}
    \end{subfigure}

    \caption{Representative one-loop realizations of the C.9
    charged-lepton texture.}
    \label{fig:C9-one-loop}
\end{figure}

% ==========================================================================
\subsection{Two-loop contributions}
% ==========================================================================

\begin{figure}[H]
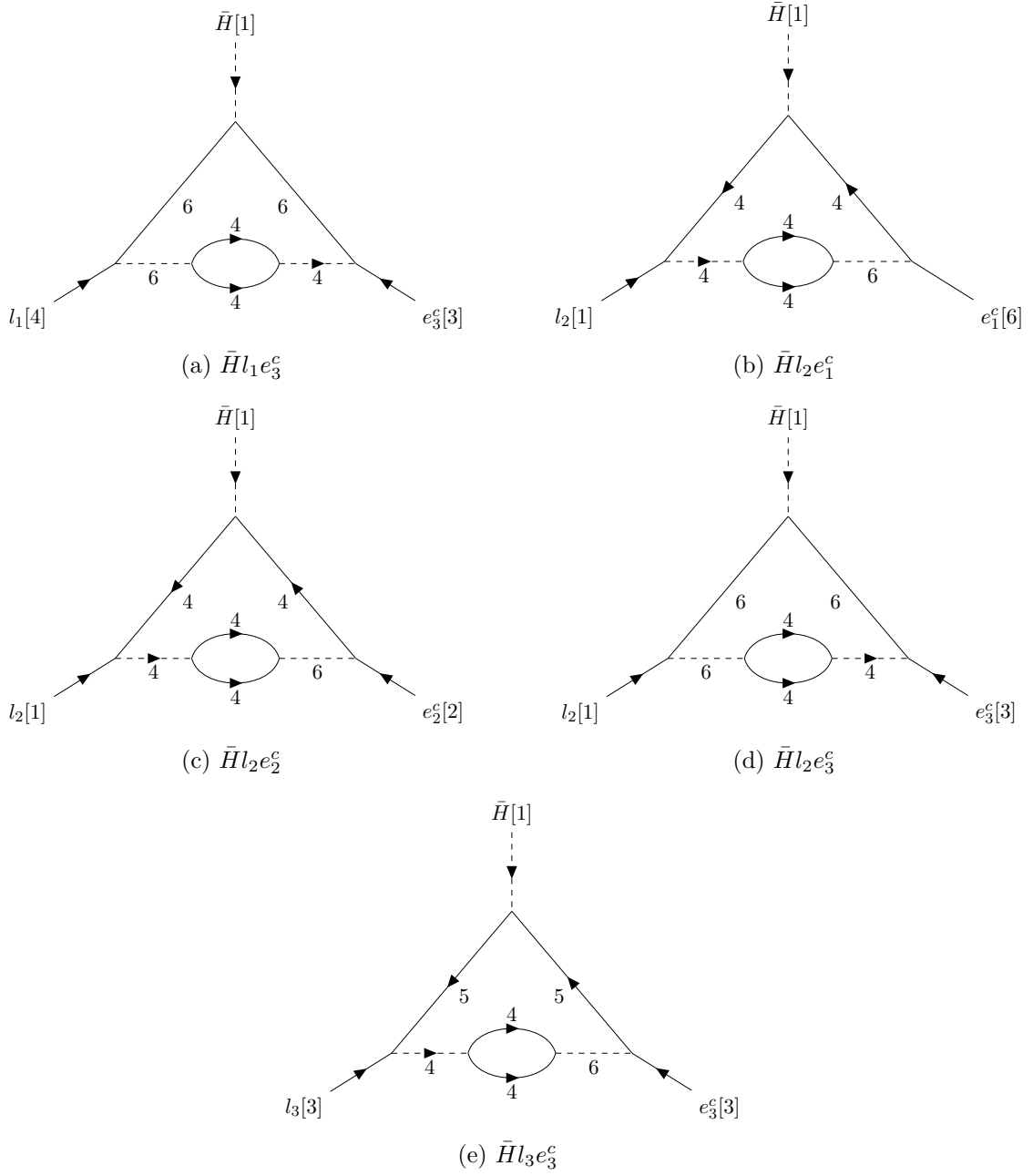

    \centering

    \begin{subfigure}[t]{0.47\textwidth}
        \centering
        \resizebox{\linewidth}{!}{
            \CNineTwoLoopDiagram
                {l_1}{4}{e_3^c}{3}
                {6}{6}{4}{6}
        }
        \caption{$\bar{H}l_1e_3^c$}
        \label{fig:C9-two-loop-l1e3}
    \end{subfigure}
    \hfill
    \begin{subfigure}[t]{0.47\textwidth}
        \centering
        \resizebox{\linewidth}{!}{%
            \CNineTwoLoopDiagram
                {l_2}{1}{e_1^c}{6}
                {4}{4}{6}{4}%
        }
        \caption{$\bar{H}l_2e_1^c$}
        \label{fig:C9-two-loop-l2e1}
    \end{subfigure}

    \medskip

    \begin{subfigure}[t]{0.47\textwidth}
        \centering
        \resizebox{\linewidth}{!}{
            \CNineTwoLoopDiagram
                {l_2}{1}{e_2^c}{2}
                {4}{4}{6}{4}
        }
        \caption{$\bar{H}l_2e_2^c$}
        \label{fig:C9-two-loop-l2e2}
    \end{subfigure}
    \hfill
    \begin{subfigure}[t]{0.47\textwidth}
        \centering
        \resizebox{\linewidth}{!}{
            \CNineTwoLoopDiagram
                {l_2}{1}{e_3^c}{3}
                {6}{6}{4}{6}
        }
        \caption{$\bar{H}l_2e_3^c$}
        \label{fig:C9-two-loop-l2e3}
    \end{subfigure}

    \medskip

    \begin{subfigure}[t]{0.47\textwidth}
        \centering
        \resizebox{\linewidth}{!}{
            \CNineTwoLoopDiagram
                {l_3}{3}{e_3^c}{3}
                {5}{4}{6}{5}
        }
        \caption{$\bar{H}l_3e_3^c$}
        \label{fig:C9-two-loop-l3e3}
    \end{subfigure}

    \caption{Representative two-loop realizations of the C.9
    charged-lepton texture. The diagrams are grouped by loop order.
    Other fusion channels may give alternative realizations of the same
    effective operators. And, for each fusion channel, we have made one choice for how to route the fermion number through the diagram.}
    \label{fig:C9-two-loop}
\end{figure}

\section{Rank-10 algebras with $\rm{cl}(\A)=3$}
\label{app:rank10}

So far we have considered fusion algebras up to rank 9 and were able to find only one with $\rm{cl}(\A)=3$, as needed to give a sufficiently broad hierarchy of textures for the up quark sector. Still, it is worth checking higher rank algebras. Ref.~\cite{anyonwiki2024,vercleyen2023lowrankfusionrings} has provided a catalogue of multiplicity free fusion rings up to rank $10$. Extending our analysis to rank $10$, we find that, among the $444$ multiplicity free algebras, there are three with a $\rm{cl}(\A)=3$ and none with $\rm{cl}(\A)>3$. They are\footnote{We suppress the multiplicity index since we are only considering multiplicity free rings. For example, $\mathrm{FR}^{10,0}_{13} \coloneqq \mathrm{FR}^{10,1,0}_{13}$ in the catalogue.} $\mathrm{FR}^{10,0}_{13}$, $\mathrm{FR}^{10,4}_{22}$
and $\mathrm{FR}^{10,8}_{11}$. This appendix subjects them to the same analysis that has been done in the body of this paper.

\subsection{The three rank 10 algebras with $\rm{cl}(\A)=3$ }

The fusion tables for each of the algebra are:

\paragraph{\href{https://anyonwiki.org/rings/10_1_0_13}{$\mathrm{FR}^{10,0}_{13}$}:}

\begin{center}
\scriptsize
\renewcommand{\arraystretch}{1.15}
\begin{tabular}{c|cccccccccc}
$\times$ & $1$ & $2$ & $3$ & $4$ & $5$ & $6$ & $7$ & $8$ & $9$ & $10$ \\
\hline
$1$ & $1$ & $2$ & $3$ & $4$ & $5$ & $6$ & $7$ & $8$ & $9$ & $10$ \\
$2$ & $2$ & $1$ & $4$ & $3$ & $6$ & $5$ & $9$ & $10$ & $7$ & $8$ \\
$3$ & $3$ & $4$ & $1$ & $2$ & $5$ & $6$ & $10$ & $9$ & $8$ & $7$ \\
$4$ & $4$ & $3$ & $2$ & $1$ & $6$ & $5$ & $8$ & $7$ & $10$ & $9$ \\
$5$ & $5$ & $6$ & $5$ & $6$ & $1+3+5$ & $2+4+6$ & $7+10$ & $8+9$ & $8+9$ & $7+10$ \\
$6$ & $6$ & $5$ & $6$ & $5$ & $2+4+6$ & $1+3+5$ & $8+9$ & $7+10$ & $7+10$ & $8+9$ \\
$7$ & $7$ & $9$ & $10$ & $8$ & $7+10$ & $8+9$ & $1+5+9+10$ & $4+6+9+10$ & $2+6+7+8$ & $3+5+7+8$ \\
$8$ & $8$ & $10$ & $9$ & $7$ & $8+9$ & $7+10$ & $4+6+9+10$ & $1+5+9+10$ & $3+5+7+8$ & $2+6+7+8$ \\
$9$ & $9$ & $7$ & $8$ & $10$ & $8+9$ & $7+10$ & $2+6+7+8$ & $3+5+7+8$ & $1+5+9+10$ & $4+6+9+10$ \\
$10$ & $10$ & $8$ & $7$ & $9$ & $7+10$ & $8+9$ & $3+5+7+8$ & $2+6+7+8$ & $4+6+9+10$ & $1+5+9+10$ \\
\end{tabular}
\end{center}
The invertible elements form a $\mathbb{Z}_2\times \mathbb{Z}_2$. The commutator filtration is
\begin{equation}
    \Com(\A)=\{1,3,5,9,10\}\subsetneq
\Com(\A)^2=A\setminus\{2\}\subsetneq\Com(\A)^3=A
\end{equation}

\paragraph{\href{https://anyonwiki.org/rings/10_1_4_22}{$\mathrm{FR}^{10,4}_{22}$}:}
\begin{center}
\scriptsize
\renewcommand{\arraystretch}{1.15}
\begin{tabular}{c|cccccccccc}
$\times$ & $1$ & $2$ & $3$ & $4$ & $5$ & $6$ & $7$ & $8$ & $9$ & $10$ \\
\hline
$1$ & $1$ & $2$ & $3$ & $4$ & $5$ & $6$ & $7$ & $8$ & $9$ & $10$ \\
$2$ & $2$ & $1$ & $4$ & $3$ & $6$ & $5$ & $9$ & $10$ & $7$ & $8$ \\
$3$ & $3$ & $4$ & $1$ & $2$ & $5$ & $6$ & $10$ & $9$ & $8$ & $7$ \\
$4$ & $4$ & $3$ & $2$ & $1$ & $6$ & $5$ & $8$ & $7$ & $10$ & $9$ \\
$5$ & $5$ & $6$ & $5$ & $6$ & $1+3+5$ & $2+4+6$ & $7+10$ & $8+9$ & $8+9$ & $7+10$ \\
$6$ & $6$ & $5$ & $6$ & $5$ & $2+4+6$ & $1+3+5$ & $8+9$ & $7+10$ & $7+10$ & $8+9$ \\
$7$ & $7$ & $9$ & $10$ & $8$ & $7+10$ & $8+9$ & $4+6+7+8$ & $1+5+7+8$ & $3+5+9+10$ & $2+6+9+10$ \\
$8$ & $8$ & $10$ & $9$ & $7$ & $8+9$ & $7+10$ & $1+5+7+8$ & $4+6+7+8$ & $2+6+9+10$ & $3+5+9+10$ \\
$9$ & $9$ & $7$ & $8$ & $10$ & $8+9$ & $7+10$ & $3+5+9+10$ & $2+6+9+10$ & $4+6+7+8$ & $1+5+7+8$ \\
$10$ & $10$ & $8$ & $7$ & $9$ & $7+10$ & $8+9$ & $2+6+9+10$ & $3+5+9+10$ & $1+5+7+8$ & $4+6+7+8$ \\
\end{tabular}
\end{center}
Similar to the previous algebra, the invertible elements here also form a $\mathbb{Z}_2\times \mathbb{Z}_2$. The commutator filtration is
\begin{equation}
    \Com(\A)=\{1,3,5,7,8\}\subsetneq
\Com(\A)^2=A\setminus\{2\}\subsetneq\Com(\A)^3=A
\end{equation}

\paragraph{\href{https://anyonwiki.org/rings/10_1_8_11}{$\mathrm{FR}^{10,8}_{11}$}:}
\begin{center}
\scriptsize
\renewcommand{\arraystretch}{1.15}
\begin{tabular}{c|cccccccccc}
$\times$ & $1$ & $2$ & $3$ & $4$ & $5$ & $6$ & $7$ & $8$ & $9$ & $10$ \\
\hline
$1$ & $1$ & $2$ & $3$ & $4$ & $5$ & $6$ & $7$ & $8$ & $9$ & $10$ \\
$2$ & $2$ & $4$ & $1$ & $5$ & $3$ & $8$ & $6$ & $10$ & $7$ & $9$ \\
$3$ & $3$ & $1$ & $5$ & $2$ & $4$ & $7$ & $9$ & $6$ & $10$ & $8$ \\
$4$ & $4$ & $5$ & $2$ & $3$ & $1$ & $10$ & $8$ & $9$ & $6$ & $7$ \\
$5$ & $5$ & $3$ & $4$ & $1$ & $2$ & $9$ & $10$ & $7$ & $8$ & $6$ \\
$6$ & $6$ & $8$ & $7$ & $10$ & $9$ & $1+7+8$ & $3+6+9$ & $2+6+10$ & $5+7+10$ & $4+8+9$ \\
$7$ & $7$ & $6$ & $9$ & $8$ & $10$ & $3+6+9$ & $5+7+10$ & $1+7+8$ & $4+8+9$ & $2+6+10$ \\
$8$ & $8$ & $10$ & $6$ & $9$ & $7$ & $2+6+10$ & $1+7+8$ & $4+8+9$ & $3+6+9$ & $5+7+10$ \\
$9$ & $9$ & $7$ & $10$ & $6$ & $8$ & $5+7+10$ & $4+8+9$ & $3+6+9$ & $2+6+10$ & $1+7+8$ \\
$10$ & $10$ & $9$ & $8$ & $7$ & $6$ & $4+8+9$ & $2+6+10$ & $5+7+10$ & $1+7+8$ & $3+6+9$ \\
\end{tabular}
\end{center}
Here, the invertible elements form a $\mathbb{Z}_5$. The commutator filtration is
\begin{equation}
    \Com(\A)=\{1,7,8\}\subsetneq
\Com(\A)^2=\{1,4,5,7,8,9,10\}\subsetneq\Com(\A)^3=A
\end{equation}

\subsection{Best assignments and textures:}
For each algebra, we scan every assignment of $(H,l_i,e_i^c,q_i,u_i^c,d_i^c)$ and minimise the median of $\rho_{\rm{tot}}$ over the ensemble, exactly as in Sec.~\ref{sec:rank8-best}. The optimum assignments we found are
\begin{equation}\begin{aligned}
\mathrm{FR}^{10,0}_{13}:&\quad H^{*}[3],\ \ell_i=(1,4,7),\ e^c_j=(1,3,9),\ q_i=(1,5,9),\ u^c_j=(2,4,5),\ d^c_j=(5,9,10)\\
\mathrm{FR}^{10,4}_{22}:&\quad H^{*}[3],\ \ell_i=(1,4,9),\ e^c_j=(1,3,8),\ q_i=(1,5,7),\ u^c_j=(2,4,5),\ d^c_j=(5,7,8)\\
\mathrm{FR}^{10,8}_{11}:&\quad H^{*}[4],\ \ell_i=(1,4,6),\ e^c_j=(2,4,8),\ q_i=(1,4,7),\ u^c_j=(1,5,7),\ d^c_j=(3,6,9)
\end{aligned}
\label{eq:rank10-best}
\end{equation}
with the resulting textures

\begin{equation}
\begin{aligned}
\mathrm{FR}^{10,0}_{13}:\qquad
Y^{e} &\sim
\begin{pmatrix}
\eps   & 1        & \eps^{2} \\
\eps^{3} & \eps^{2} & \eps^{2} \\
\eps   & \eps^{2} & \eps
\end{pmatrix},
\qquad
Y^{u} &\sim
\begin{pmatrix}
\eps^{2} & \eps^{3} & \eps \\
\eps^{2} & \eps^{2} & 1 \\
\eps     & \eps^{2} & \eps
\end{pmatrix},
\qquad
Y^{d} \sim
\begin{pmatrix}
\eps     & \eps^{2} & \eps^{2} \\
1        & \eps     & \eps \\
\eps     & \eps     & \eps^{2}
\end{pmatrix}.
\end{aligned}
\end{equation}

\begin{equation}
\begin{aligned}
\mathrm{FR}^{10,4}_{22}:\qquad
Y^{e} &\sim
\begin{pmatrix}
\eps     & 1        & \eps^{2} \\
\eps^{3} & \eps^{2} & \eps^{2} \\
\eps     & \eps^{2} & \eps
\end{pmatrix},
\qquad
Y^{u} &\sim
\begin{pmatrix}
\eps^{2} & \eps^{3} & \eps \\
\eps^{2} & \eps^{2} & 1 \\
\eps     & \eps^{2} & \eps
\end{pmatrix},
\qquad
Y^{d} \sim
\begin{pmatrix}
\eps     & \eps^{2} & \eps^{2} \\
1        & \eps     & \eps \\
\eps     & \eps^{2} & \eps
\end{pmatrix}.
\end{aligned}
\end{equation}

\begin{equation}
\begin{aligned}
\mathrm{FR}^{10,8}_{11}:\qquad
Y^{e} &\sim
\begin{pmatrix}
\eps^{2} & \eps^{3} & \eps^{2} \\
1        & \eps^{3} & \eps^{3} \\
\eps^{2} & \eps     & \eps
\end{pmatrix},
\qquad
Y^{u} &\sim
\begin{pmatrix}
\eps^{2} & \eps^{3} & \eps^{2} \\
1        & \eps^{2} & \eps \\
\eps^{2} & \eps^{3} & \eps^{2}
\end{pmatrix},
\qquad
Y^{d} \sim
\begin{pmatrix}
\eps^{3} & \eps^{2} & \eps^{3} \\
\eps^{2} & \eps     & \eps^{2} \\
\eps^{3} & \eps^{2} & \eps^{2}
\end{pmatrix}.
\end{aligned}
\end{equation}

\subsection{Scores}
The scores for the three algebras are shown in Table~\ref{tab:rank10-scores}
\begin{table}[H]
\centering
\small
\renewcommand{\arraystretch}{1.2}

\begin{tabular}{|l|c|c|c|c|c|c|}
\hline
Algebra
& $\rho_\ell$
& $\rho_u$
& $\rho_d$
& $\rho_{\rm CKM}$
& $\rho_{\rm tot}$
& $68\%$ band \\
\hline

$\mathrm{FR}^{10,0}_{13}$
& $0.602$
& $0.972$
& $0.807$
& $0.429$
& $\mathbf{0.859}$
& $[0.606,1.358]$ \\

$\mathrm{FR}^{10,4}_{22}$
& $0.602$
& $0.972$
& $0.806$
& $0.428$
& $\mathbf{0.860}$
& $[0.606,1.359]$ \\

$\mathrm{FR}^{10,8}_{11}$
& $0.601$
& $1.076$
& $0.805$
& $0.370$
& $\mathbf{0.847}$
& $[0.643,1.237]$ \\

\hline
\end{tabular}

\caption{
Median score components for the selected rank-10 fusion-ring assignments,
evaluated over the ensemble of Eq.~\eqref{eq:ensemble}.
The total score $\rho_{\rm tot}$ combines the charged-lepton, up-quark,
down-quark, and CKM contributions, while the quoted $68\%$ band gives the
central interval of the total-score distribution.
}
\label{tab:rank10-scores}
\end{table}

We note that the scores do not change appreciably from the rank 8 algebra scores. The corresponding spectra are show in Table~\ref{tab:rank10-spectra}

\begin{table}[H]
\centering
\small
\renewcommand{\arraystretch}{1.2}

\begin{tabular}{|l|c|c|c|}
\hline
Algebra
& $m_e:m_\mu:m_\tau$
& $m_u:m_c:m_t$
& $m_d:m_s:m_b$ \\
\hline

$\mathrm{FR}^{10,0}_{13}$
& $1:145.7:5411$
& $1:1957:1.16\times10^{5}$
& $1:42.35:2740$ \\

$\mathrm{FR}^{10,4}_{22}$
& $1:145.7:5411$
& $1:1957:1.16\times10^{5}$
& $1:42.30:2742$ \\

$\mathrm{FR}^{10,8}_{11}$
& $1:146.1:5414$
& $1:100.9:2.26\times10^{5}$
& $1:42.28:2745$ \\

\hline
Observed
& $1:210.7:3578$
& $1:506:1.37\times10^{5}$
& $1:19.9:1058$ \\
\hline
\end{tabular}

\caption{
Monte Carlo medians of the fermion mass ratios for the selected rank-10
fusion-ring assignments of Eq.~\eqref{eq:rank10-best}, together with the
observed ratios for comparison. The main variation among the assignments
appears in the up-type quark spectrum, particularly for
$\mathrm{FR}^{10,8}_{11}$.
}
\label{tab:rank10-spectra}
\end{table}

\subsection{The CKM matrix}
The CKM matrix across the ensemble gives the following intervals for the entries with the corresponding score,

\begin{equation}
\begin{aligned}
|V_{\rm CKM}|_{\mathrm{FR}^{10,0}_{13}}
&=
\begin{pmatrix}
0.9988^{+0.0010}_{-0.0030} &
0.0323^{+0.0441}_{-0.0228} &
0.0257^{+0.0292}_{-0.0188}
\\
0.0323^{+0.0441}_{-0.0229} &
0.9987^{+0.0010}_{-0.0030} &
0.0258^{+0.0293}_{-0.0188}
\\
0.0257^{+0.0291}_{-0.0188} &
0.0258^{+0.0294}_{-0.0187} &
0.9991^{+0.0007}_{-0.0020}
\end{pmatrix},
\qquad
\rho_{\rm CKM} &= 0.428.
\end{aligned}
\end{equation}

\begin{equation}
\begin{aligned}
|V_{\rm CKM}|_{\mathrm{FR}^{10,4}_{22}}
&=
\begin{pmatrix}
0.9988^{+0.0010}_{-0.0030} &
0.0323^{+0.0441}_{-0.0228} &
0.0259^{+0.0293}_{-0.0188}
\\
0.0323^{+0.0441}_{-0.0229} &
0.9987^{+0.0010}_{-0.0030} &
0.0258^{+0.0293}_{-0.0188}
\\
0.0259^{+0.0292}_{-0.0188} &
0.0258^{+0.0294}_{-0.0188} &
0.9991^{+0.0007}_{-0.0020}
\end{pmatrix},
\qquad
\rho_{\rm CKM} &= 0.429.
\end{aligned}
\end{equation}

\begin{equation}
\begin{aligned}
|V_{\rm CKM}|_{\mathrm{FR}^{10,8}_{11}}
&=
\begin{pmatrix}
0.7067^{+0.2051}_{-0.2958} &
0.7070^{+0.2043}_{-0.2972} &
0.0153^{+0.0225}_{-0.0110}
\\
0.7070^{+0.2043}_{-0.2972} &
0.7067^{+0.2051}_{-0.2958} &
0.0153^{+0.0224}_{-0.0110}
\\
0.0169^{+0.0205}_{-0.0093} &
0.0169^{+0.0208}_{-0.0094} &
0.9996^{+0.0003}_{-0.0011}
\end{pmatrix},
\qquad
\rho_{\rm CKM} &= 0.370.
\end{aligned}
\end{equation}

We note that the two $\mathbb{Z}_2 \times \mathbb{Z}_2$ algebras improve only slightly compared to $\mathrm{FR}_{8}^{8,6}$, whereas the algebra $\mathrm{FR}_{11}^{10,8}$ reaches a better score but it does so by rotating the 1-2 block.

\subsection{Neutrinos}
As for the neutrinos, we find that the rank 10 algebras perform worse than the rank 8 algebra. We show the results in Table~\ref{tab:rank10-nu}

\begin{table}[H]
\centering
\small
\renewcommand{\arraystretch}{1.2}

\begin{tabular}{|l|c|c|c|c|c|}
\hline
Algebra
& $m_{\nu_2}/m_{\nu_3}$
& $m_{\nu_1}/m_{\nu_3}$
& $\Delta m^2_{21}/\Delta m^2_{31}$
& $68\%$ band
& $\Lambda$ [GeV] \\
\hline

$\mathrm{FR}^{10,0}_{13}$
& $0.686$
& $0.400$
& $0.291$
& $[0.061,0.709]$
& $1.06\times10^{15}$ \\

$\mathrm{FR}^{10,4}_{22}$
& $0.584$
& $0.0125$
& $0.341$
& $[0.111,0.736]$
& $9.02\times10^{14}$ \\

$\mathrm{FR}^{10,8}_{11}$
& $0.0172$
& $0.0168$
& $9.83\times10^{-6}$
& $[2.49\times10^{-6},3.89\times10^{-5}]$
& $6.17\times10^{14}$ \\

\hline
Observed
& ---
& ---
& $3.0\times10^{-2}$
& ---
& --- \\
\hline
\end{tabular}

\caption{
Neutrino-sector predictions for the selected rank-10 fusion-ring assignments
of Eq.~\eqref{eq:rank10-best}. The scale $\Lambda$ is fixed by requiring
$m_{\nu_3}=0.05$~eV and therefore does not provide an independent discriminator
between the assignments. The observed mass-squared ratio is taken from the
NuFIT~6.0 global fit \cite{Esteban:2024eli}.
}
\label{tab:rank10-nu}
\end{table}

\section{Renormalization Group stability of the textures}
\label{app:RGstab}
In this Appendix, we show that the Yukawa textures are not ruined by renormalization group flows. Once the heavy fields are integrated out, the running of the couplings comes from loops of Standard Model fields. At one loop, as mentioned in the introduction, we get the following contributions
\begin{equation} 
\begin{split}
  \frac{dY^u}{dt}&=\frac{1}{16\pi^2}\Big[\frac{3}{2}\big(Y^uY^{u\dagger}-Y^dY^{d\dagger}\big)+\dots\Big]Y^u\\
  \frac{dY^d}{dt}&=\frac{1}{16\pi^2}\Big[\frac{3}{2}\big(Y^dY^{d\dagger}-Y^uY^{u\dagger}\big)+\dots\Big]Y^d\\
 \frac{dY^e}{dt}&=\frac{1}{16\pi^2}\Big[\frac{3}{2}Y^eY^{e\dagger}+\dots\Big]Y^e ,
  \label{eq:rge}
\end{split}
\end{equation}
where the $\dots$ represent contributions in the form of an overall multiplication factor and hence do not spoil the textures. The potentially dangerous terms are the cubics in the Yukawa couplings of the form
\begin{equation}
(\delta Y^\alpha)_{ij} \propto (Y^\beta Y^{\beta\dagger} Y^\alpha)_{ij}
\end{equation}
For the textures to be stable under RG running, we need to  show that
$$ 
(Y^\beta Y^{\beta\dagger} Y^\alpha) \leq Y^\alpha
$$
where, as in the introduction, ``$\leq$'' means that the power of $\eps$ in each entry of the matrix on the LHS is greater than or equal to the power of $\eps$ in the corresponding matrix on the RHS.

We make two restrictions on the assignment of simple objects of the fusion algebra to Standard Model fields.
\begin{itemize}
\item We assign $H$ to an invertible element of the fusion algebra (that is $1=h\cdot \overline{h}$, not merely  $1\prec h\cdot \overline{h}$).
\item To each element of the triples $q_i, u^c_i, d^c_i, l_i, e^c_i$, $i=1,2,3$ we assign a distinct element of $A$.
\end{itemize}

The terms of interest are chains of three Yukawa couplings,
\begin{equation}
  \big(Y^\alpha Y^{\alpha\dagger}Y^\beta\big)_{ij}=\sum_{k,l}Y^\alpha_{ik} Y^{\alpha*}_{lk} Y^\beta_{lj},
  \qquad x,T\in\{u,d\},
  \label{eq:chain}
\end{equation}
where $\beta$ is the sector being run and $\alpha$ the sector in the loop. With the texture inserted, one
term of the chain has size $\eps^{n^\alpha_{ik}+n^\alpha_{lk}+n^\beta_{lj}}$. Let us now define the lowest order chain as the one that has size $\eps^{s_{ij}}$ where
\begin{equation}
  s_{ij}=\min_{\alpha,k,l}\big(n^\alpha_{ik}+n^\alpha_{lk}+n^\beta_{lj}\big).
\end{equation}
We can then write the one-loop correction as, 
\begin{equation}
    \frac{\delta Y_{ij}}{Y_{ij}}\sim \eps^{s_{ij}-n_{ij}} \times \frac{1}{16\pi^2} \ln \frac{\Lambda}{\mu}
    \label{eq:correction}
\end{equation}
where we have suppressed possible order one factors (for example, coming from the top quark coupling). Since, we assume the high scale is sub-Planckian  and thus $\tfrac{1}{16\pi^2} \ln \frac{\Lambda}{\mu} < 1$, the texture survives the RG running if and only if 
\begin{equation}
    n_{ij} \leq s_{ij}\  \forall\  i,j
    \label{eq:condi}
\end{equation}
The proof is as follows. Consider first, the case $\alpha=\beta=u$. The remaining cases follow similarly. For reference, the two diagrams that contribute are,
\begin{figure}[H]
\centering
\begin{subfigure}[t]{0.48\textwidth}\centering
\resizebox{\linewidth}{!}{
\begin{tikzpicture}
  \node[circle,fill=blue,inner sep=2pt,label=below left:{\small$Y^u_{ik}$}] (a) at (0,0) {};
  \node[circle,fill=blue,inner sep=2pt,label=below right:{\small$Y^{u*}_{lk}$}] (b) at (2.6,0) {};
  \node[circle,fill=blue,inner sep=2pt,label=below right:{\small$Y^u_{lj}$}] (c) at (5.2,0) {};
  \draw[thick] (-1.3,0) node[left]{$q_i$} -- (a) -- node[above]{$u^c_k$} (b) -- node[above]{$q_l$} (c) -- (6.5,0) node[right]{$u^c_j$};
  \draw[thick,red,dashed] (a) to[out=-90,in=-90,looseness=1.3] node[below]{$H$} (b);
  \draw[thick,red,dashed] (c) -- (5.2,1.3) node[above]{$H$};
  \draw[thick,purple!70!black,dotted] (a) -- (0,1.0) node[above]{$\bar w_1$};
  \draw[thick,purple!70!black,dotted] (b) -- (2.6,1.0) node[above]{$w_2$};
  \draw[thick,purple!70!black,dotted] (c) -- (6.1,0.9) node[above]{$\bar w_3$};
\end{tikzpicture}}
\caption{Self-energy topology with chain $Y^uY^{u\dagger}Y^u$.}
\end{subfigure}\hfill
\begin{subfigure}[t]{0.48\textwidth}\centering
\resizebox{\linewidth}{!}{
\begin{tikzpicture}
  \node[circle,fill=blue,inner sep=2pt,label=above left:{\small$Y^d_{ik}$}] (a) at (0,0) {};
  \node[circle,fill=blue,inner sep=2pt,label=above left:{\small$Y^{d*}_{lk}$}] (b) at (2.6,1.4) {};
  \node[circle,fill=blue,inner sep=2pt,label=above right:{\small$Y^u_{lj}$}] (c) at (5.2,0) {};
  \draw[thick] (-1.3,0) node[left]{$q_i$} -- (a) -- node[above left]{$d^c_k$} (b) -- node[above right]{$q_l$} (c) -- (6.5,0) node[right]{$u^c_j$};
  \draw[thick,red,dashed] (a) -- node[below]{$H$ } (c);
  \draw[thick,red,dashed] (b) -- (2.6,2.7) node[above]{$H$};
  \draw[thick,purple!70!black,dotted] (a) -- (-0.6,-1.0) node[below]{$\bar w_1$};
  \draw[thick,purple!70!black,dotted] (b) -- (3.5,2.2) node[right]{$w_2$};
  \draw[thick,purple!70!black,dotted] (c) -- (5.8,-1.0) node[below]{$\bar w_3$};
\end{tikzpicture}}
\caption{Vertex topology with chain $Y^dY^{d\dagger}Y^u$.}
\end{subfigure}
\caption{The two shapes of a one-loop flavour-changing diagram. Each dot is a matched Yukawa
coupling drawn as an allowed vertex with a spurion leg.}
\label{fig:shapes}
\end{figure}
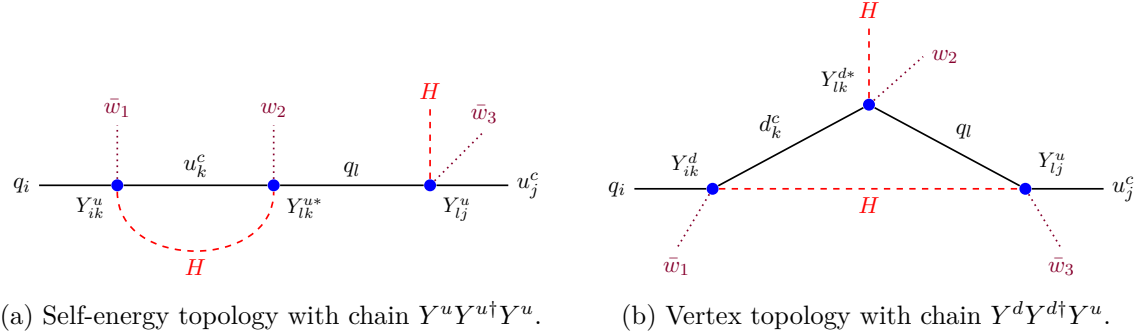

In the spurion form, we know from the vertices that 
\begin{align}
  Y^u_{ik}:&\quad 1\pr hq_iu_k \bar w_1 \qquad w_1\in\Com^{n^u_{ik}},\nonumber\\
  Y^{u*}_{lk}:&\quad 1\pr \bar h \bar q_l \bar u_k w_2 \qquad w_2\in\Com^{n^u_{lk}},\label{eq:three}\\
  Y^u_{lj}:&\quad 1\pr h q_l u_j \bar w_3 \qquad w_3\in\Com^{n^u_{lj}}\nonumber
\end{align}
But, we know that 
\begin{equation}
    1 \pr (hq_iu_k \bar w_1)(\bar h \bar q_l \bar u_k w_2)(h q_l u_j \bar w_3)
\end{equation}
Let us now use two results to simplify this relation. 
\begin{enumerate}
    \item If $X,Y$ are products of labels and $r$ a label. Then $1 \pr X\cdot r$ and $1\pr\bar r\cdot Y$ implies 
  $1 \pr X\cdot Y$. This is because, we know that $1 \pr X\cdot r$ implies $\bar r \pr X$ and similarly $r \pr Y$. Then, 
    \begin{equation}
        1 \pr r\cdot \bar r \pr X\cdot Y
    \end{equation} \\
    \item If $X $ is a product of labels and $1\pr X\cdot h\cdot \bar h$, then this means there is an element $r\pr h\cdot \bar h$ such that $1\pr X\cdot r$. But, if $h$ is assigned an invertible label, we get $r=1$ which implies $1\pr X$.
\end{enumerate}
Using the first of these, we first obtain, 
\begin{equation}
    1 \pr (hq_i \bar w_1)(\bar h w_2)(h u_j \bar w_3)
\end{equation}
Then, using the second and that $h$ is assigned an invertible label, we get 
\begin{equation}
    (\overline{hq_iu_j}) \pr (\bar w_1 w_2 \bar w_3)
\end{equation}
but note that the first paranthesis is precisely the operator we defined to be an element of $\Com^{n_{ij}}$ and $(\bar w_1 w_2 \bar w_3)\in \Com^{n^u_{ik}+n^u_{lk}+n^u_{lj}}$. Hence, we get due to filtration of the $\Com$ sets, 
\begin{eqnarray}
    n_{ij} \leq n^u_{ik}+n^u_{lk}+n^u_{lj}
\end{eqnarray}

\clearpage
\bibliographystyle{utphys}
\bibliography{ref}{}

@article{Altarelli:2010gt,
  author        = {Altarelli, Guido and Feruglio, Ferruccio},
  title         = {Discrete Flavor Symmetries and Models of Neutrino Mixing},
  eprint        = {1002.0211},
  archiveprefix = {arXiv},
  primaryclass  = {hep-ph},
  reportnumber  = {RM3-TH-10-01, CERN-PH-TH-2010-016, DFPD-10-TH-02},
  doi           = {10.1103/RevModPhys.82.2701},
  journal       = {Rev. Mod. Phys.},
  volume        = {82},
  pages         = {2701--2729},
  year          = {2010}
}

@misc{anyonwiki2024,
  title = {AnyonWiki: A Database of Fusion Categories and Anyon Models},
  url   = {https://anyonwiki.org},
  note  = {Accessed: Sep 25, 2026}
}

@article{Arkani-Hamed:1999ylh,
  author        = {Arkani-Hamed, Nima and Schmaltz, Martin},
  title         = {Hierarchies without symmetries from extra dimensions},
  eprint        = {hep-ph/9903417},
  archiveprefix = {arXiv},
  reportnumber  = {SLAC-PUB-8082},
  doi           = {10.1103/PhysRevD.61.033005},
  journal       = {Phys. Rev. D},
  volume        = {61},
  pages         = {033005},
  year          = {2000}
}

@article{Bhardwaj:2023kri,
  author        = {Bhardwaj, Lakshya and Bottini, Lea E. and Fraser-Taliente, Ludovic and Gladden, Liam and Gould, Dewi S. W. and Platschorre, Arthur and Tillim, Hannah},
  title         = {Lectures on generalized symmetries},
  eprint        = {2307.07547},
  archiveprefix = {arXiv},
  primaryclass  = {hep-th},
  doi           = {10.1016/j.physrep.2023.11.002},
  journal       = {Phys. Rept.},
  volume        = {1051},
  pages         = {1--87},
  year          = {2024}
}

@article{Choi:2021kmx,
  author        = {Choi, Yichul and Cordova, Clay and Hsin, Po-Shen and Lam, Ho Tat and Shao, Shu-Heng},
  title         = {Noninvertible duality defects in 3+1 dimensions},
  eprint        = {2111.01139},
  archiveprefix = {arXiv},
  primaryclass  = {hep-th},
  reportnumber  = {MIT-CTP/5359},
  doi           = {10.1103/PhysRevD.105.125016},
  journal       = {Phys. Rev. D},
  volume        = {105},
  number        = {12},
  pages         = {125016},
  year          = {2022}
}

@article{Choi:2022jqy,
  author        = {Choi, Yichul and Lam, Ho Tat and Shao, Shu-Heng},
  title         = {Noninvertible Global Symmetries in the {Standard Model}},
  eprint        = {2205.05086},
  archiveprefix = {arXiv},
  primaryclass  = {hep-th},
  reportnumber  = {YITP-SB-2022-21, MIT-CTP/5433},
  doi           = {10.1103/PhysRevLett.129.161601},
  journal       = {Phys. Rev. Lett.},
  volume        = {129},
  number        = {16},
  pages         = {161601},
  year          = {2022}
}

@article{Cordova:2022fhg,
  author        = {Cordova, Clay and Hong, Sungwoo and Koren, Seth and Ohmori, Kantaro},
  title         = {Neutrino Masses from Generalized Symmetry Breaking},
  eprint        = {2211.07639},
  archiveprefix = {arXiv},
  primaryclass  = {hep-ph},
  doi           = {10.1103/PhysRevX.14.031033},
  journal       = {Phys. Rev. X},
  volume        = {14},
  number        = {3},
  pages         = {031033},
  year          = {2024}
}

@article{Cordova:2022ieu,
  author        = {Cordova, Clay and Ohmori, Kantaro},
  title         = {Noninvertible Chiral Symmetry and Exponential Hierarchies},
  eprint        = {2205.06243},
  archiveprefix = {arXiv},
  primaryclass  = {hep-th},
  doi           = {10.1103/PhysRevX.13.011034},
  journal       = {Phys. Rev. X},
  volume        = {13},
  number        = {1},
  pages         = {011034},
  year          = {2023}
}

@article{Cordova:2022qtz,
  author        = {Cordova, Clay and Koren, Seth},
  title         = {Higher Flavor Symmetries in the {Standard Model}},
  eprint        = {2212.13193},
  archiveprefix = {arXiv},
  primaryclass  = {hep-ph},
  doi           = {10.1002/andp.202300031},
  journal       = {Annalen Phys.},
  volume        = {535},
  number        = {8},
  pages         = {2300031},
  year          = {2023}
}

@inproceedings{Cordova:2022ruw,
  author        = {Cordova, Clay and Dumitrescu, Thomas T. and Intriligator, Kenneth and Shao, Shu-Heng},
  title         = {{Snowmass White Paper}: Generalized Symmetries in Quantum Field Theory and Beyond},
  booktitle     = {Snowmass 2021},
  eprint        = {2205.09545},
  archiveprefix = {arXiv},
  primaryclass  = {hep-th},
  month         = {5},
  year          = {2022}
}

@article{Delaney:2026zwp,
  author        = {Delaney, Colleen and Razzouk, Ralph and Slingerland, Joost and Vercleyen, Gert},
  title         = {AnyonWiki.org},
  eprint        = {2610.30081},
  archiveprefix = {arXiv},
  primaryclass  = {math.QA},
  month         = {10},
  year          = {2026}
}

@article{Ding:2026sfm,
  author        = {Ding, Gui-Jun and King, Stephen F. and Qu, Bu-Yao},
  title         = {{Froggatt-Nielsen} mechanism with non-invertible symmetry},
  eprint        = {2610.26761},
  archiveprefix = {arXiv},
  primaryclass  = {hep-ph},
  month         = {10},
  year          = {2026}
}

@article{Dong:2025jra,
  author        = {Dong, Jun and Jeric, Tim and Kobayashi, Tatsuo and Nishida, Ryusei and Otsuka, Hajime},
  title         = {Discrete gauging and noninvertible selection rules},
  eprint        = {2507.02375},
  archiveprefix = {arXiv},
  primaryclass  = {hep-th},
  doi           = {10.1103/nsvv-l2dy},
  journal       = {Phys. Rev. D},
  volume        = {113},
  number        = {5},
  pages         = {056028},
  year          = {2026}
}

@article{Dong:2026crl,
  author        = {Dong, Jun and Kobayashi, Tatsuo and Miyamoto, Shuhei and Nishida, Ryusei and Otsuka, Hajime},
  title         = {Residual group-like symmetries in selection rules without group actions},
  eprint        = {2603.14836},
  archiveprefix = {arXiv},
  primaryclass  = {hep-th},
  reportnumber  = {EPHOU-26-02, KYUSHU-HET-356},
  month         = {3},
  year          = {2026}
}

@article{Dreiner:2008tw,
  author        = {Dreiner, Herbi K. and Haber, Howard E. and Martin, Stephen P.},
  title         = {Two-component spinor techniques and {Feynman} rules for quantum field theory and supersymmetry},
  eprint        = {0812.1594},
  archiveprefix = {arXiv},
  primaryclass  = {hep-ph},
  reportnumber  = {BN-TH-2008-12, SCIPP-08-08, FERMILAB-PUB-09-855-T},
  doi           = {10.1016/j.physrep.2010.05.002},
  journal       = {Phys. Rept.},
  volume        = {494},
  pages         = {1--196},
  year          = {2010}
}

@article{Esteban:2024eli,
  author        = {Esteban, Ivan and Gonzalez-Garcia, M. C. and Maltoni, Michele and Martinez-Soler, Ivan and Pinheiro, Jo\~ao Paulo and Schwetz, Thomas},
  title         = {{NuFit-6.0}: updated global analysis of three-flavor neutrino oscillations},
  eprint        = {2410.05380},
  archiveprefix = {arXiv},
  primaryclass  = {hep-ph},
  doi           = {10.1007/JHEP12(2024)216},
  journal       = {JHEP},
  volume        = {12},
  pages         = {216},
  year          = {2024}
}

@article{Etingof:2002vpd,
  author        = {Etingof, Pavel and Nikshych, Dmitri and Ostrik, Viktor},
  title         = {On fusion categories},
  eprint        = {math/0203060},
  archiveprefix = {arXiv},
  primaryclass  = {math.QA},
  doi           = {10.4007/annals.2005.162.581},
  journal       = {Annals Math.},
  volume        = {162},
  pages         = {581--642},
  year          = {2005}
}

@inproceedings{Feruglio:2017spp,
  author        = {Feruglio, Ferruccio},
  editor        = {Levy, Aharon and Forte, Stefano and Ridolfi, Giovanni},
  title         = {Are neutrino masses modular forms?},
  booktitle     = {{From My Vast Repertoire ...}: {Guido Altarelli's Legacy}},
  eprint        = {1706.08749},
  archiveprefix = {arXiv},
  primaryclass  = {hep-ph},
  reportnumber  = {DFPD-2017-TH-09},
  doi           = {10.1142/9789813238053_0012},
  pages         = {227--266},
  year          = {2019}
}

@article{Froggatt:1978nt,
  author       = {Froggatt, C. D. and Nielsen, Holger Bech},
  title        = {Hierarchy of Quark Masses, {Cabibbo} Angles and {CP} Violation},
  reportnumber = {CERN-TH-2519},
  doi          = {10.1016/0550-3213(79)90316-X},
  journal      = {Nucl. Phys. B},
  volume       = {147},
  pages        = {277--298},
  year         = {1979}
}

@article{Gaiotto:2014kfa,
  author        = {Gaiotto, Davide and Kapustin, Anton and Seiberg, Nathan and Willett, Brian},
  title         = {Generalized Global Symmetries},
  eprint        = {1412.5148},
  archiveprefix = {arXiv},
  primaryclass  = {hep-th},
  doi           = {10.1007/JHEP02(2015)172},
  journal       = {JHEP},
  volume        = {02},
  pages         = {172},
  year          = {2015}
}

@article{Heckman:2024obe,
  author        = {Heckman, Jonathan J. and McNamara, Jacob and Montero, Miguel and Sharon, Adar and Vafa, Cumrun and Valenzuela, Irene},
  title         = {Fate of stringy noninvertible symmetries},
  eprint        = {2402.00118},
  archiveprefix = {arXiv},
  primaryclass  = {hep-th},
  reportnumber  = {CERN-TH-2024-019},
  doi           = {10.1103/PhysRevD.110.106001},
  journal       = {Phys. Rev. D},
  volume        = {110},
  number        = {10},
  pages         = {106001},
  year          = {2024}
}

@article{Ishimori:2010au,
  author        = {Ishimori, Hajime and Kobayashi, Tatsuo and Ohki, Hiroshi and Shimizu, Yusuke and Okada, Hiroshi and Tanimoto, Morimitsu},
  title         = {Non-Abelian Discrete Symmetries in Particle Physics},
  eprint        = {1003.3552},
  archiveprefix = {arXiv},
  primaryclass  = {hep-th},
  reportnumber  = {KUNS-2260},
  doi           = {10.1143/PTPS.183.1},
  journal       = {Prog. Theor. Phys. Suppl.},
  volume        = {183},
  pages         = {1--163},
  year          = {2010}
}

@article{Kaidi:2021xfk,
  author        = {Kaidi, Justin and Ohmori, Kantaro and Zheng, Yunqin},
  title         = {{Kramers-Wannier}-like Duality Defects in (3+1)D Gauge Theories},
  eprint        = {2111.01141},
  archiveprefix = {arXiv},
  primaryclass  = {hep-th},
  doi           = {10.1103/PhysRevLett.128.111601},
  journal       = {Phys. Rev. Lett.},
  volume        = {128},
  number        = {11},
  pages         = {111601},
  year          = {2022}
}

@article{Kaidi:2024wio,
  author        = {Kaidi, Justin and Tachikawa, Yuji and Zhang, Hao Y.},
  title         = {On a class of selection rules without group actions in field theory and string theory},
  eprint        = {2402.00105},
  archiveprefix = {arXiv},
  primaryclass  = {hep-th},
  doi           = {10.21468/SciPostPhys.17.6.169},
  journal       = {SciPost Phys.},
  volume        = {17},
  number        = {6},
  pages         = {169},
  year          = {2024}
}

@article{Kaidi:2026urc,
  author        = {Kaidi, Justin},
  title         = {Introduction to Generalized Symmetries},
  eprint        = {2603.08798},
  archiveprefix = {arXiv},
  primaryclass  = {hep-th},
  month         = {3},
  year          = {2026}
}

@article{Kaplan:1991dc,
  author       = {Kaplan, David B.},
  title        = {Flavor at {SSC} energies: A New mechanism for dynamically generated fermion masses},
  reportnumber = {UCSD-PTH-91-04},
  doi          = {10.1016/S0550-3213(05)80021-5},
  journal      = {Nucl. Phys. B},
  volume       = {365},
  pages        = {259--278},
  year         = {1991}
}

@article{Kobayashi:2025cwx,
  author        = {Kobayashi, Tatsuo and Okada, Hiroshi and Otsuka, Hajime},
  title         = {Radiative neutrino mass models from non-invertible selection rules},
  eprint        = {2505.14878},
  archiveprefix = {arXiv},
  primaryclass  = {hep-ph},
  reportnumber  = {EPHOU-25-008, KYUSHU-HET-324},
  doi           = {10.1007/JHEP12(2025)111},
  journal       = {JHEP},
  volume        = {12},
  pages         = {111},
  year          = {2025}
}

@article{Kobayashi:2025lar,
  author        = {Kobayashi, Tatsuo and Mita, Hironobu and Otsuka, Hajime and Sakuma, Riku},
  title         = {Matter Symmetries in Supersymmetric {Standard Models} from Noninvertible Selection Rules},
  eprint        = {2506.10241},
  archiveprefix = {arXiv},
  primaryclass  = {hep-ph},
  reportnumber  = {EPHOU-25-009, KYUSHU-HET-325},
  doi           = {10.1093/ptep/ptag016},
  journal       = {PTEP},
  volume        = {2026},
  number        = {3},
  pages         = {033B01},
  year          = {2026}
}

@article{Kobayashi:2025ocp,
  author        = {Kobayashi, Tatsuo and Nishida, Ryusei and Otsuka, Hajime},
  title         = {Non-invertible selection rules on heterotic non-{Abelian} orbifolds},
  eprint        = {2509.10019},
  archiveprefix = {arXiv},
  primaryclass  = {hep-th},
  reportnumber  = {EPHOU-25-017, KYUSHU-HET-339},
  doi           = {10.1007/JHEP03(2026)158},
  journal       = {JHEP},
  volume        = {03},
  pages         = {158},
  year          = {2026}
}

@article{Kobayashi:2025wty,
  author        = {Kobayashi, Tatsuo and Otsuka, Hajime},
  title         = {Generalized {CP} from non-invertible selection rules},
  eprint        = {2512.16376},
  archiveprefix = {arXiv},
  primaryclass  = {hep-ph},
  reportnumber  = {EPHOU-25-019, KYUSHU-HET-345},
  month         = {12},
  year          = {2025}
}

@article{Kobayashi:2025znw,
  author        = {Kobayashi, Tatsuo and Nishioka, Yume and Otsuka, Hajime and Tanimoto, Morimitsu},
  title         = {More about quark {Yukawa} textures from selection rules without group actions},
  eprint        = {2503.09966},
  archiveprefix = {arXiv},
  primaryclass  = {hep-ph},
  reportnumber  = {EPHOU-25-003, KYUSHU-HET-312},
  doi           = {10.1007/JHEP05(2025)177},
  journal       = {JHEP},
  volume        = {05},
  pages         = {177},
  year          = {2025}
}

@article{Kobayashi:2026msq,
  author        = {Kobayashi, Tatsuo and Otsuka, Hajime and Tanimoto, Morimitsu and Yamamoto, Kei},
  title         = {Textures of dimension-six operators in the {SMEFT} with non-invertible selection rules},
  eprint        = {2608.21707},
  archiveprefix = {arXiv},
  primaryclass  = {hep-ph},
  reportnumber  = {EPHOU-26-09, KYUSHU-HET-369},
  month         = {8},
  year          = {2026}
}

@article{Leurer:1992wg,
  author        = {Leurer, Miriam and Nir, Yosef and Seiberg, Nathan},
  title         = {Mass matrix models},
  eprint        = {hep-ph/9212278},
  archiveprefix = {arXiv},
  reportnumber  = {RU-92-59, WIS-92-94-PH},
  doi           = {10.1016/0550-3213(93)90112-3},
  journal       = {Nucl. Phys. B},
  volume        = {398},
  pages         = {319--342},
  year          = {1993}
}

@article{Leurer:1993gy,
  author        = {Leurer, Miriam and Nir, Yosef and Seiberg, Nathan},
  title         = {Mass matrix models: The Sequel},
  eprint        = {hep-ph/9310320},
  archiveprefix = {arXiv},
  reportnumber  = {RU-93-43, WIS-93-93-PH},
  doi           = {10.1016/0550-3213(94)90074-4},
  journal       = {Nucl. Phys. B},
  volume        = {420},
  pages         = {468--504},
  year          = {1994}
}

@article{McGreevy:2022oyu,
  author        = {McGreevy, John},
  title         = {Generalized Symmetries in Condensed Matter},
  eprint        = {2204.03045},
  archiveprefix = {arXiv},
  primaryclass  = {cond-mat.str-el},
  doi           = {10.1146/annurev-conmatphys-040721-021029},
  journal       = {Ann. Rev. Condensed Matter Phys.},
  volume        = {14},
  pages         = {57--82},
  year          = {2023}
}

@article{Otsuka:2026fpy,
  author        = {Otsuka, Hajime and Yokokura, Ryo},
  title         = {Non-Invertible {Froggatt-Nielsen} Symmetry},
  eprint        = {2610.21688},
  archiveprefix = {arXiv},
  primaryclass  = {hep-ph},
  reportnumber  = {KYUSHU-HET-377},
  month         = {10},
  year          = {2026}
}

@article{ParticleDataGroup:2024cfk,
  author        = {Navas, S. and others},
  collaboration = {Particle Data Group},
  title         = {Review of particle physics},
  doi           = {10.1103/PhysRevD.110.030001},
  journal       = {Phys. Rev. D},
  volume        = {110},
  number        = {3},
  pages         = {030001},
  year          = {2024}
}

@article{Schafer-Nameki:2023jdn,
  author        = {Schafer-Nameki, Sakura},
  title         = {{ICTP} lectures on (non-)invertible generalized symmetries},
  eprint        = {2305.18296},
  archiveprefix = {arXiv},
  primaryclass  = {hep-th},
  doi           = {10.1016/j.physrep.2024.01.007},
  journal       = {Phys. Rept.},
  volume        = {1063},
  pages         = {1--55},
  year          = {2024}
}

@inproceedings{Shao:2023gho,
  author        = {Shao, Shu-Heng},
  title         = {What's Done Cannot Be Undone: {TASI} Lectures on Non-Invertible Symmetries},
  booktitle     = {{Theoretical Advanced Study Institute in Elementary Particle Physics 2023}: {Aspects of Symmetry}},
  eprint        = {2308.00747},
  archiveprefix = {arXiv},
  primaryclass  = {hep-th},
  reportnumber  = {YITP-SB-2023-19},
  month         = {8},
  year          = {2023}
}

@article{Suzuki:2025bxg,
  author        = {Suzuki, Motoo and Xu, Ling-Xiao and Zhang, Hao Y.},
  title         = {Spurion Analysis for Non-Invertible Selection Rules from Near-Group Fusions},
  eprint        = {2508.14970},
  archiveprefix = {arXiv},
  primaryclass  = {hep-ph},
  month         = {8},
  year          = {2025}
}

@article{Suzuki:2025kxz,
  author        = {Suzuki, Motoo and Xu, Ling-Xiao},
  title         = {Spurion analysis of {$\mathbb{Z}_{M}/\mathbb{Z}_{2}$} non-invertible selection rules: low-order versus all-order zeros},
  eprint        = {2510.18972},
  archiveprefix = {arXiv},
  primaryclass  = {hep-ph},
  doi           = {10.1007/JHEP02(2026)227},
  journal       = {JHEP},
  volume        = {02},
  pages         = {227},
  year          = {2026}
}

@article{Tambara:1998vmj,
  author  = {Tambara, Daisuke and Yamagami, Shigeru},
  title   = {Tensor Categories with Fusion Rules of Self-Duality for Finite Abelian Groups},
  doi     = {10.1006/jabr.1998.7558},
  journal = {J. Algebra},
  volume  = {209},
  number  = {2},
  pages   = {692--707},
  year    = {1998}
}

@article{vercleyen2023lowrankfusionrings,
  author        = {Vercleyen, Gert and Slingerland, Joost K.},
  title         = {On low rank fusion rings},
  eprint        = {2205.15637},
  archiveprefix = {arXiv},
  primaryclass  = {math-ph},
  doi           = {10.1063/5.0148848},
  journal       = {J. Math. Phys.},
  volume        = {64},
  number        = {9},
  pages         = {091703},
  year          = {2023}
}

\end{document}